\documentclass[10pt]{article}

\usepackage{amsmath}
\usepackage{amssymb,dsfont} 

\usepackage{graphicx,colordvi} 

\usepackage{pstool}
\usepackage{standalone}
\usepackage{booktabs}
\usepackage{lineno}
\usepackage{float}
\usepackage{microtype}

\usepackage[normalem]{ulem}

\newcommand{\parenths}[1]{\left( #1 \right)}
\newcommand{\brackets}[1]{\left[ #1 \right]}
\newcommand{\braces}[1]{\left\{ #1 \right\}}
\newcommand{\abs}[1]{\left| #1 \right|}

\newcommand{\ev}[1]{\langle #1 \rangle}
\newcommand{\Ev}[1]{\Bigl\langle #1 \Bigr\rangle}

\newcommand{\argmax}{\operatornamewithlimits{argmax}}
\newcommand{\argmin}{\operatornamewithlimits{argmin}}

\newcommand{\pd}[2]{\frac{\partial #1}{\partial #2}}
\newcommand{\spd}[3]{\frac{\partial ^2 #1}{\partial #2 \partial #3}}

\newcommand{\Dt}[0]{\Delta t}
\newcommand{\tr}[0]{\mathrm{tr}}

\newcommand{\diag}[0]{\mathrm{diag}}

\newcommand{\usom}{u_{\mathrm{som}}}
\newcommand{\usomi}[1]{u_{\mathrm{som},{#1}}}

\newcommand{\com}[1]{{\vspace{5mm} \noindent\Large\color{red}#1\color{black}\vspace{5mm}}}

\newcommand{\soutt}[1]{\sout{#1} }
\newcommand{\temporary}[1]{#1}

\renewcommand{\soutt}[1]{}   
\renewcommand{\com}[1]{} 
\renewcommand{\temporary}[1]{} 

\usepackage[labelfont=bf,labelsep=period,justification=raggedright]{caption}

\makeatletter
\renewcommand{\@biblabel}[1]{\quad#1.}
\makeatother

\date{}

\begin{document}

\begin{flushleft}
{\temporary{\Huge Highlighted Changes \\ }
\Large
\textbf{A statistical model for \emph{in vivo} neuronal dynamics}
}
\\
\vspace{0.5cm}
Simone Carlo Surace$^{1,2,\ast}$, 
Jean-Pascal Pfister$^{2}$
\\
\vspace{0.5cm}
\bf{1} Department of Physiology, University of Bern, Bern, Switzerland
\\
\bf{2} Institute of Neuroinformatics, University of Zurich and ETH Zurich, Zurich, Switzerland
\\
$\ast$ E-mail: surace@ini.uzh.ch
\end{flushleft}

\temporary{
\color{blue} Blue: New content in revised manuscript. \color{black}

\sout{Crossed out:} Content present in the original manuscript but removed in the revision.

\color{red} Red: Entire section moved or removed. \color{black}
\\

Layout changes from PLOS Computational Biology to PLOS One are not marked explicitly.
}
\section*{Abstract}

Single neuron models have a long tradition in computational neuroscience. Detailed biophysical models such as the Hodgkin-Huxley model as well as simplified neuron models such as the class of integrate-and-fire models relate the input current to the membrane potential of the neuron. Those types of models have been extensively fitted to \emph{in vitro} data where the input current is controlled. Those models are however of little use when it comes to characterize intracellular \emph{in vivo} recordings since the input to the neuron is not known. Here we propose a novel single neuron model that characterizes the statistical properties of \emph{in vivo} recordings. More specifically, we propose a stochastic process where the subthreshold membrane potential follows a Gaussian process and the spike emission intensity depends nonlinearly on the membrane potential as well as the spiking history. We first show that the model has a rich dynamical repertoire since it can capture arbitrary subthreshold autocovariance functions, firing-rate adaptations as well as arbitrary shapes of the action potential. We then show that this model can be efficiently fitted to data without overfitting. Finally, we show that this model can be used to characterize and therefore precisely compare various intracellular \emph{in vivo} recordings from different animals and experimental conditions.

\section{Introduction}
During the last decade, there has been an increasing number of studies providing intracellular \emph{in vivo} recordings. From the first intracellular recordings performed in awake cats \cite{Woody:1978tm,Baranyi:1993tb} to more recent recording in cats \cite{Steriade:2001vh}, monkeys \cite{Matsumura:1988uq}, mice \cite{Poulet:2008kb}, and even in freely behaving rats \cite{Lee:2006ej}, it has been shown that the membrane potential displays large fluctuations and is very rarely at the resting potential. Some recent findings in the cat visual cortex have also suggested that the statistical properties of spontaneous activity is comparable to the neuronal dynamics when the animal is exposed to natural images \cite{ElBoustani:2009jo}. Similar results have been found in extracellular recordings in the ferret \cite{Berkes:2011ga}. Those data are typically characterized by simple quantifications such as the firing rate or the mean subthreshold membrane potential \cite{Poulet:2008kb}, but a more comprehensive quantification is often missing. So the increasing amount of intracellular data of awake animals as well as the need to compare in a rigorous way the data under various recording conditions call for a model of spontaneous activity in single neurons.

Single neuron models have been studied for more than a century. Simple models such as the integrate-and-fire model \cite{Lapicque:1907wh,Stein:1967er} 
and its more recent nonlinear versions \cite{Latham:2000us,FourcaudTrocme:2003wz,Brette:2005ke} describe the relationship between the input current and the membrane potential in terms of a small number of parameters and are therefore convenient for analytical treatment, but do not provide much insight about the underlying biophysical processes. On the other end of the spectrum, biophysical models such as the Hodgkin-Huxley model \cite{Hille:2001cv,Hodgkin:1952td} relate the input current to the membrane potential through a detailed description of the various transmembrane ion channels, but estimating the model parameters remains challenging \cite{Gerstner:2009fp,Druckmann:2007exa}. Despite the success of those types of models, none of them can be directly applied to intracellular \emph{in vivo} recordings for the simple reason that the input current is not known.

Another reason why a precise model of spontaneous activity is needed is that there are several theories that have been proposed that critically depend on statistical properties of spontaneous activity. For example Berkes et al. validate their Bayesian treatment of the visual system by comparing the spontaneous activity and the averaged evoked activity \cite{Berkes:2011ga}. Another Bayesian theory proposed the idea that short-term plasticity acts as a Bayesian estimator of the presynaptic membrane potential \cite{Pfister:2010kf}. To validate this theory, it is also necessary to characterize spontaneous activity with a statistical model that describes the subthreshold as well as the spiking dynamics.

The last motivation for a model that describes both the subthreshold and the suprathreshold dynamics is the possibility to separate those two dynamics in a principled way. Indeed, it is interesting to know from the recordings what reflects the input dynamics and what aspect comes from the neuron itself (or rather what is associated with the spiking dynamics). Of course a simple voltage threshold can separate the sub- and a suprathreshold dynamics, but the value of the threshold is somewhat arbitrary and could lead to undesirable artifacts. Therefore a computationally sound model that decides itself what belongs to the subthreshold and what belongs to the suprathreshold dynamics is highly desirable. 

Here, we propose a single neuron model that describes intracellular \emph{in vivo} recordings as a sum of a sub- and suprathreshold dynamics. This model is flexible enough in order to capture the large diversity of neuronal dynamics while remaining tractable, i.e. the model can be fitted to data in a reasonable time. More precisely, we propose a stochastic process where the subthreshold membrane potential follows a Gaussian process and the firing intensity is expressed as a non-linear function of the membrane potential. Since we further include refractoriness and adaptation mechanisms, our model, which we call the Adaptive Gaussian Point Emission process (AGAPE), can be seen as an extension of both the log Gaussian Cox process \cite{Moller:1998tx} and the generalized linear model \cite{Truccolo:2005hz,Pillow:2008bo,Paninski:2009fz}.

\section{Results}

\begin{figure}[htbp]
\centering
\includegraphics[width=\textwidth]{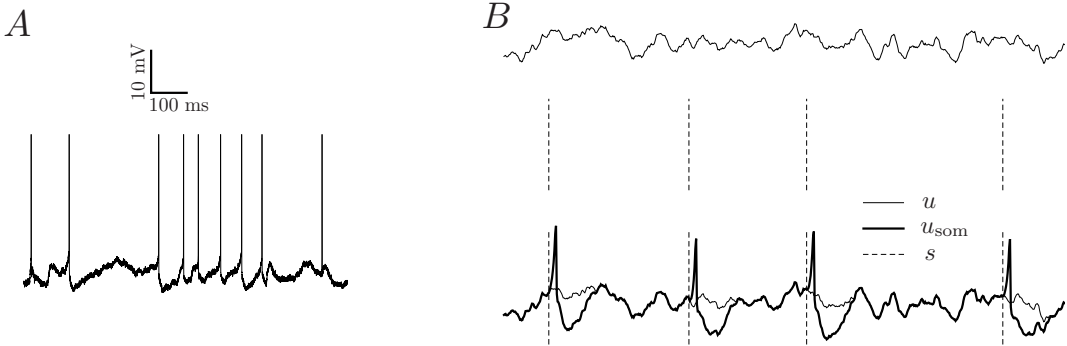}
\caption{(A) A sample \emph{in vivo} membrane potential trace from an intracellular recording of a neuron in HVC of a Zebra Finch. (B) The generative AGAPE model can generate a trace of subthreshold membrane potential $u$ (top trace). Based on this potential, a spike train $s$ is generated (middle, dashed vertical lines). Finally, a stereotypic spike-related kernel is convolved with the spike train and added to $u$, giving rise to $u_{\text{som}}$ (bottom, thick line). This quantity is the synthetic analog of the recorded, preprocessed \emph{in vivo} membrane potential.}
\label{Fig:1}
\end{figure}

Here we present a statistical model of the subthreshold membrane potential and firing pattern of a single neuron \emph{in vivo}. See Fig.~\ref{Fig:1}A for such an \emph{in vivo} membrane potential recording. We first provide a formal definition of the model and then show a range of different results. 1) The model is flexible and supports arbitrary autocorrelation structures and adaptation kernels. Therefore, the range of possible statistical features is very large. 2) The model is efficiently fittable and the learning procedure is validated on synthetic data. 3) The model can be fitted to \emph{in vivo} datasets. 4) All the features included in the model are required to provide a good description of in vivo data.

\subsection{Definition of the AGAPE model}

The AGAPE model is a single neuron model where the input to the neuron is not known, which is typically the case under \emph{in vivo} conditions. The acronym AGAPE stands for Adaptive GAussian Point Emission process since the subthreshold membrane potential follows a Gaussian process and since the spike emission process is adaptive. 

More formally, the AGAPE process defines a probability distribution p($\usom,s)$ over the somatic membrane voltage trace $\usom(t)$ and the spike train $s(t)=\sum_{i=1}^{n_s}\delta(t-\hat t_i)$ where $\hat t_i$, $i=1,...,n_s$ are the nominal spike times (decision times), which occur a certain fixed time period $\delta>0$ before the peak of the action potential. From this probability distribution (or generative model) it is possible to draw samples that look like intracellular \emph{in vivo} activity (for practical purposes, the samples will be compared to the preprocessed recordings, see explanations below). The AGAPE model assumes that the somatic membrane voltage as a function of time $\usom(t)$ is given by (see Fig.~\ref{Fig:1})
\begin{equation}
\usom(t)=u_r+u(t)+u_{\mathrm{spike}}(t),
\label{eqn:uraw}
\end{equation}
where $u_r$ is a constant (the reference potential), $u(t)$ describes the subthreshold membrane potential as a stochastic function drawn from a stationary Gaussian process (GP) \cite{Rasmussen:2006tj}
\begin{equation}
\label{eqn:u}
u\sim\mathcal{GP}\brackets{0,k(t-t')}
\end{equation}
with covariance function $k(t-t')$ (which can be parametrized by a weighted sum of exponential decays with weights $\sigma^2_i$ and inverse time constants $\theta_i$, see Materials and Methods). For small values of $\delta$ (e.g. 1-3 ms), $u(t)$ can be seen as the net contribution from the unobserved synaptic inputs and $u_{\mathrm{spike}}(t)$ is the spike-related contribution (see Fig.~\ref{Fig:1}B) which consists of the causal convolution of a stereotypical spike-related kernel $\alpha$ with the spike train $s(t)$, i.e.
\begin{equation}
u_{\mathrm{spike}}(t)=\int_{0}^{\infty}\alpha(t')s(t-t')dt'.
\label{eqn:uspike}
\end{equation}
where $\alpha$ can be parametrized by a weighted sum of basis functions with weights $a_i$, see Materials and Methods. Here, we have made a separation of subthreshold and suprathreshold layers, in that whatever is stereotypic and triggered by the point-like spikes $s(t)$ is attributed to $u_{\mathrm{spike}}(t)$, and the rest belongs to the fluctuating signal $u(t)$. This separation need not correspond to the biophysical distinction between synaptic inputs and active processes of the recorded cell (i.e. the positive feedback loop of the spiking mechanism). Indeed, especially for a choice of large $\delta$ (e.g. $\sim$ 20 ms), $u_{\mathrm{spike}}(t)$ also contains large depolarizations due to strong synaptic input which cannot be explained by the GP signal $u(t)$. 

Note that this model could easily be extended by including an additional term in Eq.~(1) which depends on an external input, e.g. a linear filter of the input (see also Discussion). However, since this input current was not accessible in our recordings, its contribution was assumed to be part of $u(t)$ or $u_{\mathrm{spike}}(t)$. 

Now we proceed to the coupling between the subthreshold potential $u(t)$ and the spiking output, as well as adaptive effects associated with spike generation. These effects are summarized by an instantaneous firing rate $r(t)$ -- as in the generalized linear model (GLM) \cite{Truccolo:2005hz,Pillow:2008bo,Paninski:2009fz} or escape-rate models \cite{Gerstner:2002usa} -- which is computed from the value of the subthreshold membrane potential at time $t$, $u(t)$, and the spike history as
\begin{equation}
r(t)=g\brackets{A(t)+\beta u(t)}, \quad A(t)=\int_0^{\infty}\eta(t') s(t-t')dt',
\label{eqn:gf}
\end{equation}
where $\beta\geq0$ is the coupling strength between $u$ and the spikes, and $A(t)$ is the adaptation variable which is the convolution of an adaptation kernel $\eta$ (which can be parametrized by a weighted sum of basis functions with weights $w_i$, see Materials and Methods) with the past spike train. Also note that we choose not to model adaptation currents explicitly, since they would simultaneously impact the membrane potential and the firing probability (see Discussion). The function $g$ is called gain function, and here we use an exponential one, i.e. $g\brackets{A(t)+\beta u(t)}=e^{\log r_0+A(t)+\beta u(t)}$. Other functional forms such as rectified linear or sigmoidal could be used depending on the structure of the data. However, this choice has important implications on the efficiency of learning of the model parameters \cite{Paninski:2004gi}. We define the probability density for $s$ on an interval $[0,T]$ conditioned on $u$ as
\begin{equation}
\label{eqn:s|u}
\begin{split}
p(s|u) &= \exp\parenths{-\int_0^Tr(t)\ dt}\prod_{i=1}^{n_s} r(\hat t_i),\quad s(t)=\sum_{i=1}^{n_s}\delta(t-\hat t_i), \quad 0\leq \hat t_1 < ... < \hat t_{n_s}\leq T,\quad n_s\in \mathds{N}_0.
\end{split}
\end{equation}
The parameter $\beta$ connects the subthreshold membrane potential $u$ to the rate fluctuations. The magnitude of the rate fluctuations depend on the variance $\sigma^2$ of $u$, and therefore we use $\beta\sigma$ as a measure of the effective coupling strength. When $\beta>0$ the quantity $\theta(t)=-A(t)/\beta$ can be regarded as a soft threshold variable which is modulated after a spike, and $u(t)-\theta(t)$ is the effective membrane potential relevant for the spike generation. This spiking process is a point process which generalizes the log Gaussian Cox process. Indeed, when $A=0$, Eq.~\eqref{eqn:s|u} describes an inhomogeneous Poisson process with rate $g\brackets{\beta u(t)}$. 

Practically, if we want to draw a sample from the AGAPE process, we first draw a sample $u$ from the Gaussian Process (see S1 Text for how to do this efficiently), then for each time $t$ we draw spikes $s(t)$ with probability density $r(t)$ and update the adaptation variable $A(t)$. Finally, the somatic membrane potential is calculated using Eq.~\ref{eqn:uraw}. 

It is important to emphasize at this point that while the model may be directly fitted to the raw membrane potential $u_{\mathrm{raw}}$ as recorded by an intracellular electrode, we median filter the data in order to avoid artifacts and downsample for computational efficiency (see `Materials and Methods'). In this study the model is always fitted to the preprocessed recordings $u^{\ast}_{\text{som}}$ and this is reflected e.g. in the shape of $\alpha$ which is most strongly affected by the preprocessing. It is important to keep in mind this point while interpreting the results of model fitting. The details of the preprocessing steps which were used are given in the `Materials and Methods' section.

\subsection{The model has a rich dynamical repertoire}
The AGAPE provides a flexible framework which can be adjusted in complexity to model a wide range of dynamics. While for the datasets presented here a covariance function was used which consists of a sum of Ornstein-Uhlenbeck (OU) kernels, the Gaussian Process (GP) allows for arbitrary covariance functions to be used. This includes simple exponential decay (as produced by a leaky integrate-and-fire neuron driven by white noise current), but it can produce also more interesting covariance functions such as power-law covariances, which are reported in \cite{Pozzorini:2013bj, ElBoustani:2009jo}, or subthreshold oscillations, as reported in \cite{Buzsaki:2002ui}.

\begin{figure}[htbp]
\centering
\includegraphics[width=\textwidth]{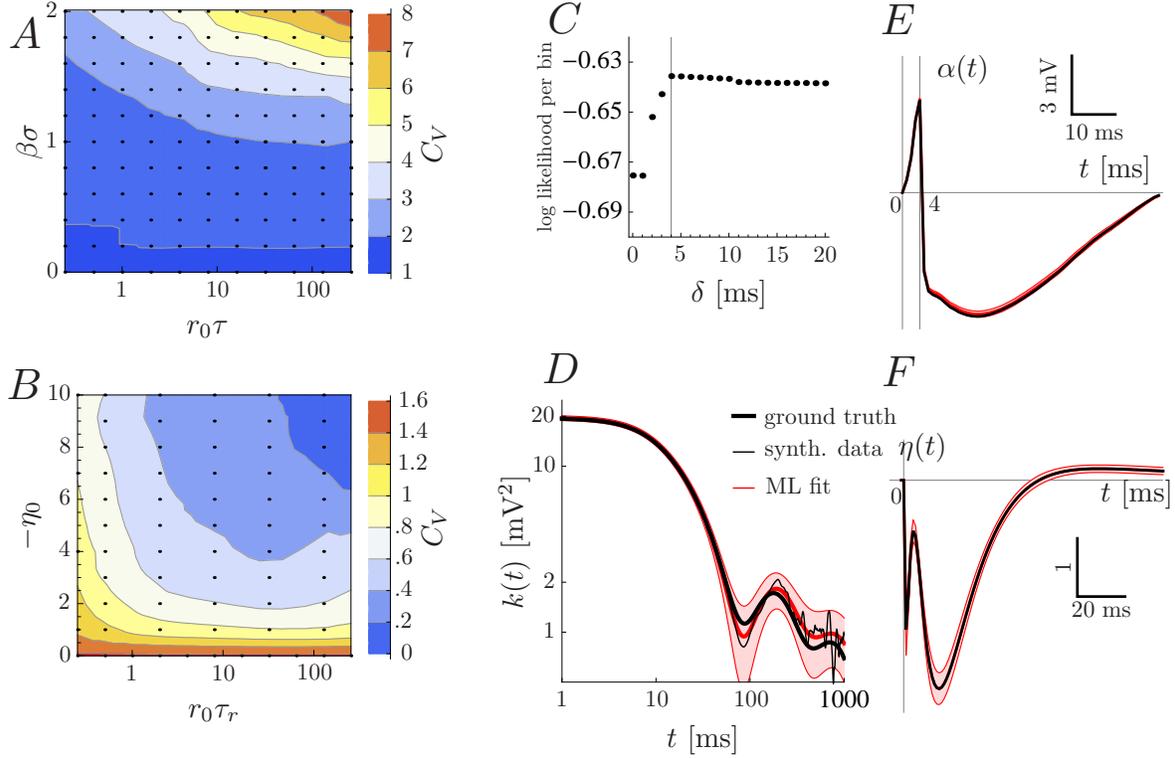}
\caption{The model has a rich dynamical repertoire (A,B) and can be correctly fitted to synthetic data (C-F). (A,B) The coefficient of variation ($C_V$) of the inter-spike interval distribution is computed for parameter values shown as black dots and then linearly interpolated. (A) The $C_V$ of a simple version of the AGAPE ($k(t)=\sigma^2 e^{-t/\tau}$, $\alpha=\eta=0$) as a function of the model parameters (membrane time-constant $\tau$, baseline firing rate $r_0$ and coupling strength $\beta\sigma$). (B) $C_V$ of the AGAPE model with an exponentially adaptive process with fixed membrane time-constant, firing rate and coupling ($\beta\sigma=r_0\tau=1$) as a function of the parameters describing adaptation, namely adaptation strength $\eta_0$ and time-constant $\tau_r$. (C,D,E,F) Synthetic data is sampled from the AGAPE model with GP (D), spike-related (E), and adaptation (F) kernels as depicted in black, and $\delta=4$ ms, $r_0=4.15$ Hz, $\beta=0.374$ mV$^{-1}$. Then the AGAPE is fitted to the synthetic data by maximum likelihood (ML). (C) The maximum log likelihood per bin as a function of the parameter $\delta$ has its maximum at the ground truth value $\delta=4$ ms. (D,E,F) The ML estimates (red) of the GP, spike-related and adaptation kernels lie within two standard deviations (red shaded regions, estimated by means of the observed Fisher information) from the ground truth.} 
\label{Fig:2}
\end{figure}

The model is also able to reproduce a wide range of firing statistics. A common measure of firing irregularity is the coefficient of variation ($C_V$, i.e. the ratio of standard deviation and mean) of the inter-spike interval distribution. In the absence of adaptation, the AGAPE is a Cox process and therefore has a coefficient of variation $C_V\geq 1$ \cite{Shinomoto:2001if}. The precise value of the $C_V$ is a function of the coupling strength ($\beta\sigma$) as well as the autocorrelation of the GP. To illustrate this, we sampled synthetic data from a simple version of the AGAPE where the subthreshold potential $u$ is an OU process with time-constant $\tau$. As shown in Fig.~\ref{Fig:2}A, the $C_V$ is an increasing function of the membrane time-constant $\tau$, baseline firing rate $r_0$, and dimensionless coupling parameter between membrane potential and firing rate $\beta\sigma$. Moreover, the range of the $C_V$ extends from $1$ to $\approx$ 8 within a range of $\beta\sigma\in[0,2]$ and $r_0\tau\in[2^{-2},2^{8}]$. In the presence of adaptation, firing statistics are markedly different and can produce values of $C_V<1$ \cite{Gerstner:2002usa,Lindner:2002gb}. To illustrate this point, we considered an exponential adaptation kernel, i.e. $\eta(t)=-\eta_0 e^{-t/\tau_{r}}$. While the $C_V$ increases as a function of $\beta\sigma$ and $r_0\tau$ as before, the range of values of the $C_V$ now also covers the interval $(0,1)$ which is not accessible by the Cox process but which is observed in many neurons across the brain \cite{Softky:1993uj}. In order to study the influence of the parameters of the adaptation mechanism, we fix $\beta\sigma=r_0\tau=1$ and plot the $C_V$ as a function of $r_0\tau_{r}$ and $\eta_0$ (see Fig.~\ref{Fig:2}B). Within the parameter region explored in Fig.~\ref{Fig:2}B, the $C_V$ spans values from 0.1 up to 1.6.

\subsection{The model can be learned efficiently}
The parameters of the AGAPE model are learned through a maximum likelihood approach. More precisely, we fit the model to an \emph{in vivo} sample (highlighted by a `$\ast$') of preprocessed somatic membrane potential $u^{\ast}_{\text{som}}$ and spike train $s^{\ast,\delta}$ by maximizing the log likelihood applied to the joint data set $(u^{\ast}_{\text{som}},s^{\ast,\delta})$ over the parameter space of the model (i.e. $u_r,\log r_0,\beta$, the coefficients of the kernels $k$, $\eta$, and $\alpha$, and the delay parameter $\delta$). The empirical spike train $s^{\ast,\delta}$ depends on the parameter $\delta$ because the formal spike times $\hat{t}_i$ are assigned to be a time period $\delta$ before the recorded peak of the action potential. The joint probability of the data can be expressed as a product
\begin{equation}
\begin{split}
p(u_{\text{som}}^{\ast},s^{\ast,\delta})&=\int p(u) p(s^{\ast,\delta}|u) p(u_{\text{som}}^{\ast} |u,s^{\ast,\delta}) \mathcal{D}u\\
		&=\int p(u) p(s^{\ast,\delta}|u)\delta(u_{\text{som}}^{\ast}-u-u_r-u_{\mathrm{spike}}) \mathcal{D}u\\
		&=p(u=u_{\text{som}}^{\ast}-u_r-u_{\mathrm{spike}}) p(s^{\ast,\delta}|u=u_{\text{som}}^{\ast}-u_r-u_{\mathrm{spike}})\\
		&\equiv p_{s^{\ast,\delta}}(u_{\text{som}}^{\ast}) p(s^{\ast,\delta}|u_{\text{som}}^{\ast}).
\end{split}
\label{eqn:p}
\end{equation}
The subscript $s^{\ast,\delta}$ of the first factor denotes the explicit dependence on the spike train. The individual terms on the r.h.s. will be given below. The function we are optimizing is the logarithm of the above joint probability which we can write as
\begin{equation}
\mathcal{L}(u_r, k, \alpha, \log r_0, \beta, \eta, \delta) = \log p_{s^{\ast,\delta}}(u^{\ast}_{\text{som}} ; u_r, k, \alpha) + \log p(s^{\ast,\delta} | u^{\ast}_{\text{som}}; u_r, \alpha, \log r_0, \beta, \eta).
\end{equation}
It should be noted that the presence of the spike-related kernel $\alpha$ in both terms produces a trade-off situation: removing the spike-related trajectory improves the Gaussianity of the membrane potential $u$ (and therefore boosts the first term) at the cost of the of the second term by removing the short upward fluctuation that leads to the spike. This trade-off situation makes maximum likelihood parameter estimation a non-concave optimization problem. Moreover, the evaluation of the GP likelihood of $n$ samples, where $n=\mathcal{O}(10^5)$, comes at a high computational cost. Two important techniques make the parameter learning both tractable and fast: the first is the use of the circulant approximation of the GP covariance matrix which makes the evaluation of the likelihood function fast. The second is the use of an alternating fitting algorithm which (under an appropriate parametrization, see `Materials and Methods') replaces the non-concave optimization in the full parameter space with two concave optimizations and a non-concave one in suitable parameter subspaces. Those two techniques are further described in the next section.

\subsubsection{Efficient likelihood computation}
The log-likelihood function is evaluated in its discrete-time form with $n$ time points separated by a time-step $\Dt$. The GP variable $u$ (which leads to $\usom$ through Eq.~\eqref{eqn:uraw}) is multivariate Gaussian distributed with a covariance matrix $K_{ij}=k(t_i-t_j)$, where $t_i=i\Dt$. The matrix $K$ is symmetric and, by virtue of stationarity, Toeplitz. Evaluation of the GP likelihood requires inversion of $K$, which is computationally expensive (the time required to invert a matrix typically scales with $n^3$). For this reason we approximate this Toeplitz matrix by the circulant matrix $C$ which minimizes the Kullback-Leibler divergence (see \cite{Katsaggelos:1991ex,Bach:2004dw,Gray:2006ta} and S1 Text\ref{SI:1})
\begin{equation}
C=\argmin_{D\; \mathrm{circulant}} \mathcal{D}_{\mathrm{KL}}\brackets{\mathcal{N}(m,K)||\mathcal{N}(m,D)}
\label{eqn:circmin}
\end{equation}
between the two multivariate Gaussian distributions with the same mean but different covariance matrices. This optimization problem can be solved by calculating the derivative of $\mathcal{D}_{\mathrm{KL}}\brackets{\mathcal{N}(m,K)||\mathcal{N}(m,D)}$ with respect to $D$ and using the diagonalization of $D$ by a Fourier transform matrix \cite{Gray:2006ta}. After a bit of algebra (see S1 Text\ref{SI:1}), denoting $k_i=K_{1i}$ and $k_{n+1}\equiv 0$, the optimal circulant matrix can be written as $C_{ij}=c_{(i-j\!\!\mod n) +1}$, where $i,j=1,...,n$ and
\begin{equation}
c_i=\frac{1}{n}\brackets{(n-i+1)k_i+(i-1)k_{n-i+2}}.
\end{equation}
The replacement of $K$ by $C$ is equivalent to having periodic boundary conditions on $u$, which has a small effect under the assumption that the time interval spanned by the data is much longer than the largest temporal autocorrelation length of $k$. So the first term on the r.h.s. of Eq.~\eqref{eqn:p} is a multivariate Gaussian density $\mathcal{N}(0,C)$. The determinant of the covariance matrix $C$ is the product of eigenvalues, which for a circulant matrix are conveniently given by the entries of $\hat{c}$, the discrete Fourier transform of $c$ (see the S1 Text\ref{SI:1} for our conventions regarding discrete Fourier transforms). Also the scalar product $u^TC^{-1}u$ can be written in terms of $\hat{c}$. Together, the first term on the r.h.s. of Eq.~\eqref{eqn:p} takes the simple form
\begin{equation}
\log p_{s^{\ast,\delta}}(\usom^{\ast})=-\frac{1}{2}\sum_{i=1}^n\parenths{\log (2\pi\hat c_i)+\frac{1}{n}\frac{\abs{\hat u_i}^2}{\hat c_i}},
\label{eqn:psofucirc}
\end{equation}
where $\hat{u}_i$ are the components of the discrete Fourier transform of $u^{\ast}$. The Gaussian component of the membrane potential $u$ is implicitly given by the discretized somatic voltage modified by a discrete-time version of the spike-related kernel convolution,
\begin{equation}
u^{\ast}_i=u_{\text{som},i}^{\ast}-u_r-\sum_{j=1}^{i-1} \alpha_j s^{\ast,\delta}_{i-j},
\label{eqn:udiscrete}
\end{equation}
where $s^{\ast,\delta}_{i}$ is the binned spike train (see below), $\alpha_i$ is a discretized version of the spike-related kernel. The time required to compute $\log p_s(\usom^{\ast})$ is determined by the complexity of the Fourier transform, which is of the order of $n\log n$. This dramatic reduction in complexity (compared to $n^3$) allows a fast evaluation of the log-likelihood.	

The spiking distribution $p(s^{\ast,\delta}|\usom^{\ast})$ is approximated by a Poisson distribution with constant rate within one time bin. For each bin, $s^{\ast,\delta}_i$ counts the number of spikes that occur in that bin, and the conditional likelihood of the spikes therefore reads
\begin{equation}
\log p(s^{\ast,\delta}|\usom^{\ast})=\sum_{i=1}^{n} \braces{s^{\ast,\delta}_i \log \brackets{r_i\Dt}-r_i\Dt-\log \brackets{s^{\ast,\delta}_i!}},
\label{eqn:psgvuraw}
\end{equation}
where $r_i = g[\beta u^{\ast}_i+\sum_{j=1}^{i-1} \eta_j s^{\ast,\delta}_{i-j}]$ and $u^{\ast}_i$ as defined in Eq.~\eqref{eqn:udiscrete}. If $s^{\ast,\delta}_i$ contains only zeros and ones (which can be accomplished given small enough bins), the last term $\log s^{\ast,\delta}_i!$ vanishes.

\subsubsection{Efficient parameter estimation}

\begin{figure}[htbp]
\centering
\includegraphics[width=0.8\textwidth]{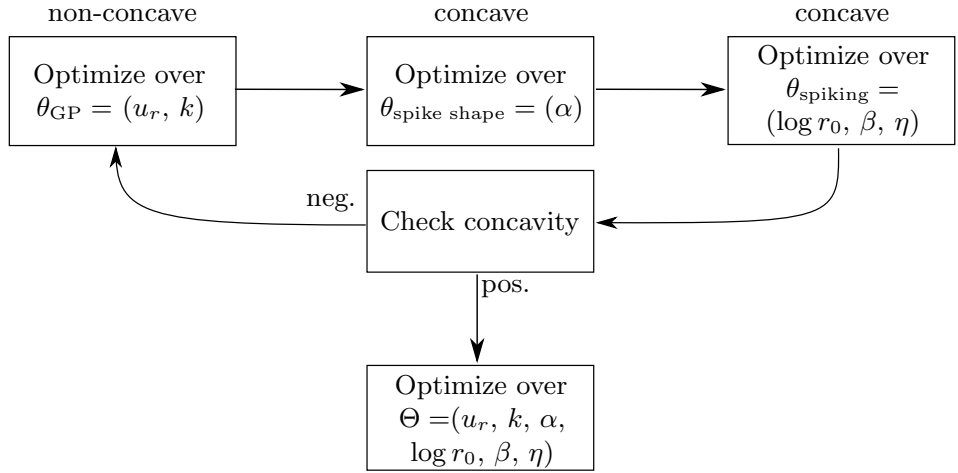}
\caption{This schematic shows the optimization scheme that is used to learn the parameters of the AGAPE model when it is fitted to the data (for a given $\delta$). As long as the current parameter estimate sits in a non-concave region of the likelihood function, the top cycle optimizes over different subspaces of the parameter space. If and when a concave point is reached, the optimization proceeds in the full parameter space. This whole scheme is repeated for each value of $\delta$ in order to find the optimal one. }
\label{Fig:3}
\end{figure}

Except for the parameter $\delta$, which takes discrete values of multiples of the discretization step $\Dt$, it is possible to analytically calculate the first and second partial derivatives of the likelihood function defined in Eq.~\eqref{eqn:p} with respect to the model parameters $(u_r, k, \alpha, \log r_0, \beta, \eta)$ (see S1 Text\ref{SI:1}) to facilitate the use of gradient ascent, Newton, and pseudo-Newton optimization algorithms. A desirable feature of an optimization problem is concavity of the objective function (in our case, the log-likelihood function). Even though the problem of finding optimal parameters for the AGAPE process is not concave, the optimization can be done in three alternating subspaces (see Fig.~\ref{Fig:3}). The full set of parameters $\Theta$ is divided into three parts: $\theta_{\text{GP}}$ for the GP parameters ($u_r$, parameters of $k$), $\theta_{\text{spike kernel}}$ for the spike-related kernel parameters, and $\theta_{\text{spiking}}$ for the parameters controlling spike emission ($\log r_0,\beta$ and parameters of $\eta$). The optimization then proceeds according to the following cycle: (1) the GP parameters are learned, (2) the spike-related kernel parameters are learned, and lastly (3) the spiking parameters are learned. In each step the remaining parameters are held fixed. The cycle is repeated until the parameters reach a region where the log likelihood is locally concave in the full parameter space, after which the optimization can be run in the full parameter space until it converges. Joint concavity of the log likelihood holds if all the eigenvalues of the Hessian matrix are strictly negative. As shown in \cite{Paninski:2004gi}, step (3) is concave for a certain class of gain functions $g$, including the exponential function, and linear parametrizations of the adaptation kernel. The same holds for the spiking term of the log-likelihood in step (2). The voltage term of the log likelihood of step (2) is concave by numerical inspection in the cases we considered. To summarize, steps (2) and (3) are concave and Newton's method can be used in these steps as well as for the final concave optimization in the full space. Step (1) is non-concave and therefore a simple gradient ascent algorithm is used. 

The optimization over $(u_r, k, \alpha, \log r_0, \beta, \eta)$ is repeated for every $\delta=0,\Dt,2\Dt,$~$...,\delta_{\text{max}}$ in order to select the one $\delta$ that maximizes the log-likelihood $\mathcal{L}(u_r, k, \alpha, \log r_0, \beta, \eta, \delta)$. The value of $\delta_{\text{max}}$ is chosen such that it is less than the least upper bound of the support of the basis of the spike-related kernel $\alpha$. Since the parameters $u_r, k, \alpha, \log r_0, \beta, \eta$ are expected to change only a little when going from one $\delta$ to the next, $\delta+\Dt$, learned parameters for $\delta$ can be used as initial guesses for nearby $\delta+\Dt$ or $\delta-\Dt$. We thus get two different initializations, which we can exploit by starting e.g. with $\delta=0$, ascending through the sequence of candidate $\delta$'s up to the maximum $\delta$, and descending back to zero.

\subsection{Validation with synthetic data}

Despite this improvement in speed and tractability, the optimization is still riddled with multiple local minima which require the use of multiple random initializations. In order to demonstrate the validity of the fitting method, synthetic data of length 270.112 seconds ($n=270112$, the same as \emph{in vivo} dataset $\mathcal{D}_1$, see below) was generated with known parameters ($\delta=4$ ms, $r_0=4.15$ Hz, $\beta=0.374$ mV$^{-1}$ and GP, spike-related kernel and adaptation kernels as depicted in Fig.~\ref{Fig:2}D-F). The learning algorithm was initialized with least-squares estimates of the covariance function parameters $\sigma^2_i$ based on the empirical autocorrelation function of $\usom$ and spike-related kernel and adaptation kernels set to zero. The true underlying $\delta$ can be recovered from the synthetic data (Fig.~\ref{Fig:2}C). Moreover, the algorithm converges after a few dozen iterations (taking only three minutes on an ordinary portable computer) and -- with $\delta$ set to 4 ms -- recovers the correct GP, spike-related, and adaptation kernels (Fig.~\ref{Fig:2}D-F). All ML estimates lie within a region of two standard deviations around the ground truth, where standard deviations are estimated from the observed Fisher information \cite{Efron:1978gm}.

\subsection{The model can fit \emph{in vivo} data} 
We fitted the model to a number of \emph{in vivo} traces from different animals and conditions (see `Materials and Methods' for a detailed description of the data sets). We would like to remind the reader at this point that the model is never fitted to the raw membrane potential, but to a preprocessed, i.e. median-filtered and downsampled dataset (see Materials and Methods). Because of this preprocessing stage, the model only sees the truncated action potentials which emerge from the median filter. This is reflected in the extracted spike-related kernel $\alpha$, which is characterized by a smaller amplitude than the original action potential in the raw membrane potential data.

\begin{figure}[htbp]
\centering
\includegraphics[width=\textwidth]{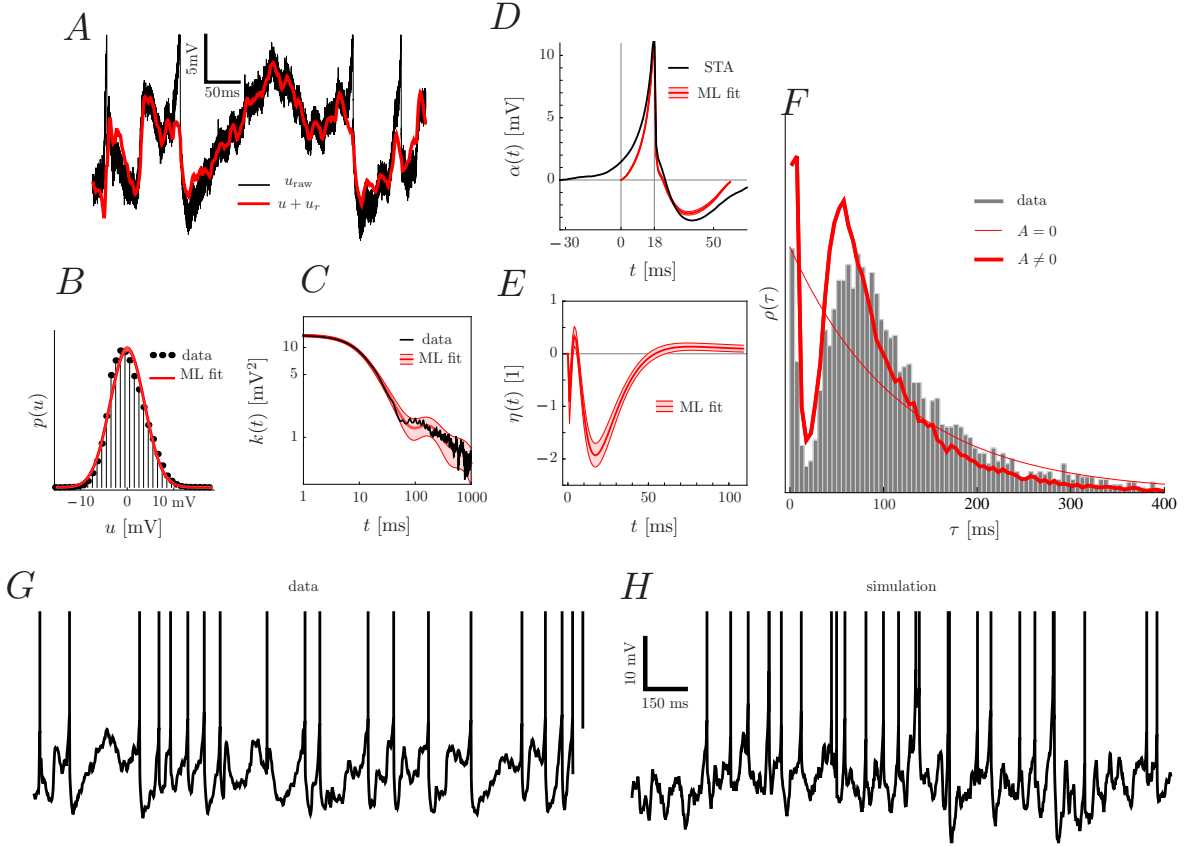}
\caption{The results of maximum likelihood (ML) parameter fitting to dataset $\mathcal{D}_1$. After fitting, we see (A) the removal of the spike-related kernel through the difference between the recorded trace $u^{\ast}_{\text{som}}$ and the subthreshold membrane potential $u+u_r$; (B) the match of the stationary distribution of the subthreshold potential $u$ and a Gaussian. We also observe that (C) the autocorrelation function of the data, Eq.~\eqref{eqn:kemp}, is well reproduced by $k(t)$ in Eq.~\eqref{eqn:k}; (D) the spike-related kernel $\alpha(t)$ starts at $-\delta=-18$ ms relative to the peak of the action potential. The difference between the spike-triggered average (STA) and the spike-related kernel is attributed to the GP; and (E) that the adaptation kernel $\eta(t)$ shows distinct modulation of firing rate which produces firing statistics significantly different from a Poisson process. This is also reflected in the inter-spike interval density $\rho(\tau)$ (F) of the data, which shows good qualitative agreement with a simulated AGAPE with adaptive kernel as in (E) (thick red line), but not by a non-adaptive (i.e. Poisson) process (thin red line). After fitting, a two second sample of synthetic data (H) looks similar as the \emph{in vivo} data (G). In (G,H) vertical lines are drawn at the spiking times. All red shaded regions denote $\pm$ 2 standard deviations, estimated from the observed Fisher information. }
\label{Fig:4}
\end{figure}

We show the detailed results of the model fitting for the example songbird HVC dataset $\mathcal{D}_1$. The optimal value of $\delta$ for this dataset was $\delta=18$ ms (see S3 Fig), with which the model captures the subthreshold and suprathreshold statistics (smaller values of $\delta$ compromise both the subthreshold and suprathreshold description because the large upward fluctuations which preceed spikes in this dataset are unlikely to arise from a GP). In particular, the stationary distribution of the membrane potential $u$ is well approximated by a Gaussian (Fig.~\ref{Fig:4}B) and pronounced after-hyperpolarization is seen in the spike-related kernel (Fig.~\ref{Fig:4}D). The subthreshold autocorrelation structure is well reproduced by the parametric autocorrelation function $k$ (Fig.~\ref{Fig:4}C). The adaptation kernel reveals an interesting structure in the way the spiking statistics deviates from a Poisson process (Fig.~\ref{Fig:4}E). This feature of the spiking statistics is also reflected in the inter-spike interval (ISI) distribution (Fig.~\ref{Fig:4}F). Both the data and the fitted model first show an increased, and then a significantly decreased probability density when compared to a pure Poisson process. The remaining parameters are listed in Tab.~\ref{Tab:1} (errors denote two standard deviations, estimated from Fisher information, see Materials and Methods). The model can be used to generate synthetic data, which is shown in Fig.~\ref{Fig:4}H.

\begin{figure}[htbp]
\centering
\includegraphics[width=0.8\textwidth]{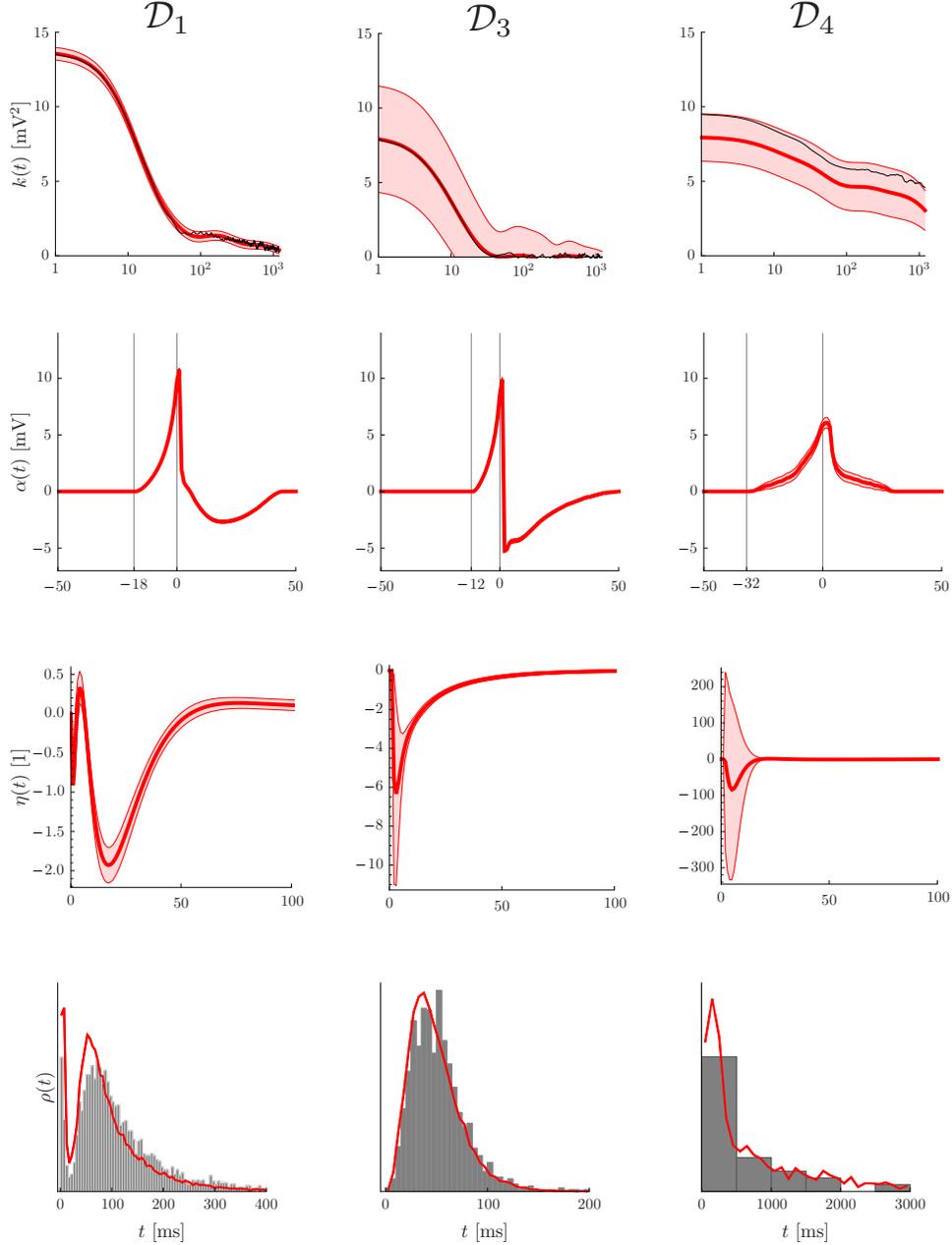}
\caption{Fitting results for three different datasets. Dataset $\mathcal{D}_1$ is the same as in Fig.~\ref{Fig:4}, i.e. an HVC neuron from anesthetized Zebra Finch. $\mathcal{D}_3$ is from HVC in awake Zebra Finch, and $\mathcal{D}_4$ is from mouse visual cortex in awake mouse. The different panels show the results after fitting; in the first line the GP covariance function $k(t)$ (red) and the empirical autocorrelation (black), Eq.~\eqref{eqn:kemp}, in the second line the spike-related kernel $\alpha(t)$, in the third line the adaptation kernel $\eta(t)$, and in the fourth line the inter-spike interval density $\rho(t)$ (data ISI histogram in gray, simulated ISI distribution from AGAPE in red). There are pronounced differences between datasets in all three kernels, showing the flexibility of the AGAPE model in describing a wide range of statistics. All red shaded regions denote $\pm$ 2 standard deviations, estimated from the observed Fisher information.
}
\label{Fig:5}
\end{figure}

In order to show the generality of the model, we fitted the model on two more datasets, $\mathcal{D}_3$ from another HVC neuron and $\mathcal{D}_4$ from mouse visual cortex. The parameter $\delta$ was found to take the optimal value of 12 ms for $\mathcal{D}_3$ and 32 ms for $\mathcal{D}_4$ (to see how fitted parameters change as a function of $\delta$, see S4 Fig and S5 Fig). The comparison of the GP, spike-related and adaptation kernels is shown in Fig.~\ref{Fig:5}, and the remaining parameters are listed in Tab.~\ref{Tab:1}. The three cells show pronounced differences in autocorrelation structure, spike-related kernel and spike-history effects. In particular the two datasets $\mathcal{D}_1$ and $\mathcal{D}_4$ show rather long autocorrelation lengths of the membrane potential and asymmetric spike-related kernels, whereas the cell in $\mathcal{D}_3$ has comparatively short autocorrelation length and very pronounced hyperpolarization. Adaptation is much stronger in $\mathcal{D}_3$ than in $\mathcal{D}_1$, balancing the much higher baseline firing rate $r_0$, see Tab.~\ref{Tab:1}. The error bars on the adaptation kernel are small for datasets $\mathcal{D}_1$ and $\mathcal{D}_3$ due to the abundance of spikes. On the other hand, the adaptation kernel of dataset $\mathcal{D}_4$ is poorly constrained by the available data. This is due to the fact that dataset $\mathcal{D}_4$ consists of very short trials with very few spikes. Despite this fact, good agreement is achieved between the distribution of inter-spike intervals of the \emph{in vivo} data and ISI statistics sampled from the AGAPE (see Fig.~\ref{Fig:5}, bottom row) for all datasets.

\begin{table}
\centering
\begin{tabular}{@{} *4l @{}}    \toprule
Dataset & $\mathcal{D}_1$ & $\mathcal{D}_3$ & $\mathcal{D}_4$  \\\midrule
$\delta$ [ms] 		& 18  				& 12  				& 32   				\\ 
$u_r$ [mV] 			& -52.9$\pm$0.2 	& -66.6$\pm$0.1 	& -51.5$\pm$0.5	\\ 
$r_0$ [Hz] 			& 11.7$\pm$0.6 	& 71$\pm$5 		& 0.15$\pm$0.05	\\
$\beta$ [mV$^{-1}$] & 0.12$\pm$0.01	& 0.24$\pm$0.02	& 0.46$\pm$0.05	\\\midrule
$\beta \sigma$ [1] 	& 0.45$\pm$0.03 	& 0.67$\pm$0.04	& 1.3$\pm$0.1		\\\bottomrule
 \hline
\end{tabular}
\caption{The values (p.m. two standard deviations, estimated from the observed Fisher information) of the fitted parameters not shown in Fig.~\ref{Fig:5} for the \emph{in vivo} datasets described in the main text. The last row shows the effective coupling strength between the membrane potential and the firing rate, given by $\beta$ times the standard deviation $\sigma$ of the membrane potential.}
 \label{Tab:1}
\end{table}

\subsection{The model does not overfit \emph{in vivo} data}

The AGAPE process has a fairly large number of parameters. Therefore it is important to check whether the model overfits the data, compromising its generalization performance. In short, when a model has too many parameters, it tends to be poorly constrained by the data and therefore when the model is first trained on one part of the data and then tested on another part on which it is not trained, the test performance will be significantly worse than the training performance. 

Here, we use cross-validation to perform a factorial model comparison on an exemplary dataset in order to validate the different structural parts of the model. The procedure is described in detail in the Materials and Methods.

\begin{figure}[htbp]
\centering
\includegraphics[width=0.6\textwidth]{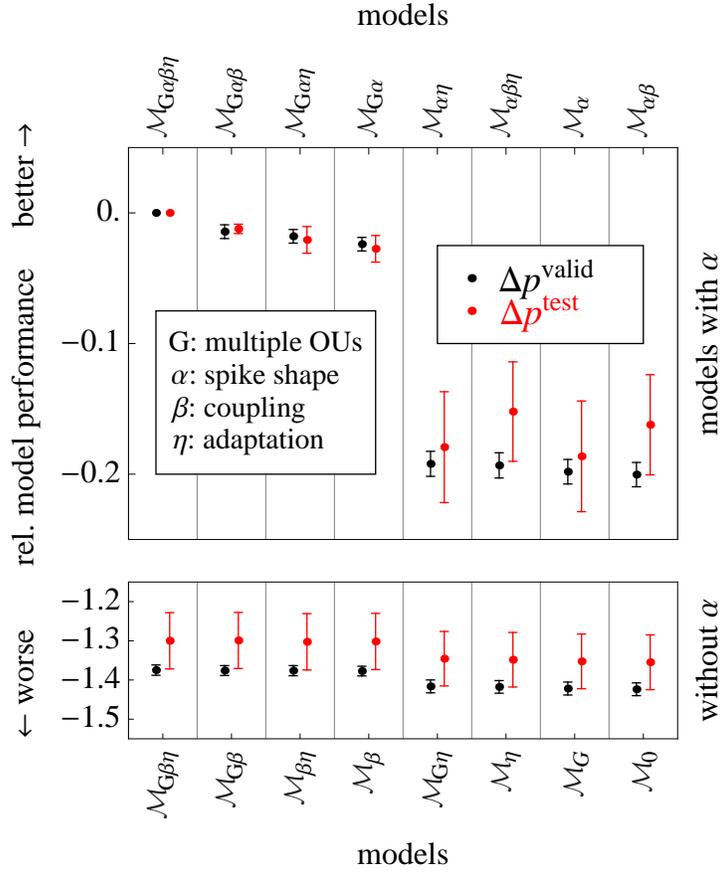}
\caption{Comparison of the different models on dataset $\mathcal{D}_2$. The relative measure of model performance, i.e. the per-bin log-likelihood $\Delta p$ (see Eq.~\eqref{eqn:deltap}) between any model and the most complex model ($\mathcal{M}_{G\alpha\beta\eta}$) are significantly negative (with exception of $\mathcal{M}_{G\alpha\eta}$, and trivially $\mathcal{M}_{G\alpha\beta\eta}$) , implying that the added complexity improves the model fit without overfitting. This holds for both validation scores $\Delta p^{\mathrm{valid}}$ (black) and scores from unseen test data $\Delta p^{\mathrm{test}}$ (red). Error bars denote one standard error of the mean (S.E.M.). The biggest improvement of fit quality is achieved by including the spike-related kernel (upper vs. lower part of the figure).}
\label{Fig:6}
\end{figure}

Model comparison is performed on the dataset $\mathcal{D}_2$ and the results are shown in Fig.~\ref{Fig:6}, where the mean difference of per-bin log-likelihood (see `Materials and Methods')
\begin{equation}
\Delta p^{\mathrm{valid}}_i=\ev{\Delta p^{\mathrm{valid}}_{ij}}_j, \quad \Delta p^{\mathrm{test}}_i=\ev{\Delta p^{\mathrm{test}}_{ij}}_j, \qquad \Delta p^{\mathrm{valid,test}}_{ij}=p^{\mathrm{valid,test}}_{i,j}-p^{\mathrm{valid,test}}_{G\alpha\beta\eta,j}
\label{eqn:deltap}
\end{equation}
is shown for all models $i\in\braces{0,...,G\alpha\beta\eta}$ (here, $\ev{\cdot}_j$ denote averages over chunks $j$ of the cross-validation). The results are very similar for both validation data (which was left-out during training, but appeared in other training runs) and the test data which was never seen during training. The most complex model ($\mathcal{M}_{G\alpha\beta\eta}$) performs significantly better than any one of the simpler models on validation data except $\mathcal{M}_{G\alpha\eta}$ where the difference is too small and lies inside a region of two standard errors of the mean. This confirms that most of the model features are required to provide an accurate description of the experimental data. 

\section{Discussion}

In this study, we introduced the AGAPE generative model for single-neuron statistics in order to describe the spontaneous dynamics of the somatic potential without reference to an input current. We showed that this model has a rich dynamical repertoire and can be fitted to data efficiently. By fitting a heterogeneous set of data, we finally demonstrated that the AGAPE model can be used for the systematic characterization and comparison of in vivo intracellular recordings. 

\subsection{Flexibility and tractability of the model}
The AGAPE model provides a unified description of intracellular dynamics, offering a large degree of flexibility in accounting for the distinct statistical features of a neuron. As the example datasets demonstrate, the model readily teases apart the differences in the statistics which exist between different cells in different animals (see Fig.~\ref{Fig:5}). This shows that the model is sensitive enough to distinguish between datasets which are in fact very similar.

We used a set of approximations and techniques to make the model fitting tractable, despite the non-concavity of the log likelihood function. It is still the case that multiple local maxima of the likelihood function can make the fitting somewhat hard, especially if the quantity of data available for fitting is quite low. However, since one run of the fitting itself takes only a few minutes even on a portable computer, multiple initializations can be tried out in a relatively short amount of time.

\subsection{Comparison with existing models}
From an operational perspective, existing spiking neuron models can be divided into three main categories: stimulus-driven, current-driven and input-free spiking neurons. The first category contains phenomenological models that relate sensory stimuli to the spiking output of the neuron. The linear-nonlinear-Poisson model (LNP) \cite{Chichilnisky:2001ua}, the generalized linear model (GLM) \cite{Truccolo:2005hz,Pillow:2008bo,Paninski:2009fz} or the GLM with additional latent variables \cite{Vidne:2012is} are typical examples in this category. Even though the spike generation of the AGAPE shares some similarities with those models, there is an important distinction to make. In those models the convolved input (i.e. the output of the `L' step of the LNP or the input filter of the GLM) is an internal variable that does not need to be mapped to the somatic membrane potential whereas in our case, the detailed modeling of the membrane potential dynamics is an important part of the AGAPE. Consequently, those phenomenological models are descriptions of extracellular spiking recordings whereas the AGAPE models the dynamics of the full membrane potential accessible with intracellular methods. 

The second class of spiking models aims at bridging the gap between the input current and the spiking output. The rather simple integrate-and-fire types of models such as the exponential integrate-and-fire \cite{Brette:2005ke} or the spike-response model \cite{Gerstner:2002usa,Jolivet:2006dp} as well as the more biophysical models such as the Hodgkin-Huxley model \cite{Hodgkin:1952td} fall within this category. In contrast to those models where the action potentials are caused by the input current, the AGAPE produces a fluctuating membrane potential and stochastic spikes without a reference to an input current. 

The last category of models aims at producing spontaneous spiking activity without an explicit dependence to a given input \cite{Cunningham:2007wl,Pfister:2010kf,Macke:2011ut}. For example, Cunningham et al. propose a doubly stochastic process where the spiking generation is given by a gamma interval process and the firing intensity by a rectified Gaussian process, which provides a flexible description of the firing statistics \cite{Cunningham:2007wl}. However, the membrane potential dynamics is not modeled. In opposition, the neuronal dynamics assumed by Pfister et al. \cite{Pfister:2010kf} models explicitly the membrane potential (as a simple Ornstein-Uhlenbeck process) but is not flexible enough to capture the dynamics of \emph{in vivo} recordings. Also any of the current-driven spiking neuron models mentioned above can be turned into an input-independent model by assuming some additional input noise. So why is there a need to go beyond stochastic versions of those models? An integrate-and-fire model with additive Gaussian white noise is certainly fittable, but does not have the flexibility to model arbitrary autocorrelation for the membrane potential. At the other end of the spectrum, a Hodgkin-Huxley model with some colored noise would certainly be able to model a richer dynamical repertoire, but the fitting of it remains challenging \cite{Gerstner:2009fp} (but see \cite{Druckmann:2007exa}). The main advantage of the AGAPE is that it is at the same time very flexible and easily fittable. The flexibility mostly comes from the fact that any covariance function can be assumed for the GP process. The relative ease of fitting comes from the circulant approximation as well as from the presence of concave subspaces in the full parameter space.

Another distinct feature of our model with respect to other existing models is the explicit modeling of the spike-related trajectory instead of the spike-triggered average (as e.g. in \cite{Mensi:2012fu}). Even though both concepts share similarities - both would capture a sudden and strong input that lead to a spike - there is an important distinction. The spike-triggered average also captures the (possibly smaller) upward fluctuations of the membrane potential which causes the spike while the spike-related kernel $\alpha$ precisely avoids capturing those fluctuations, letting the GP kernel explain them.

So if we removed the spike-triggered average e.g. in synthetic data where the true coupling parameter $\beta$ is large, we would also remove the characteristic upward fluctuation of the membrane potential which causes the spike. By doing so, the fitting procedure would not find the correct relation between the values of the membrane potential and the observed spike patterns and therefore choose a $\beta$ close to zero. Thus, if something has to be removed around an action potential (and our model comparison, Fig.~\ref{Fig:6}, demonstrates convincingly that this is necessary), the formulation of the model demands that it is parametrically adjustable. This is the main reason why in our model framework the spike-triggered average has to be rejected as a viable extraction method. Note that if the true coupling parameter $\beta$ is close to zero, the spike-triggered average is close to the extracted spike-related kernel $\alpha$. For data where the action potential shape shows considerable variability, the model could be generalized to include a stochastic or a history-dependent spike-related kernel.

\subsection{Extensions and future directions} 
Despite the focus of the present work on single-neuron spontaneous dynamics, the AGAPE model admits a straightforward inclusion of both stimulus-driven input and recurrent input. The inclusion of stimulus-driven input is similar as for the GLM model and allows the model to capture the neuronal correlate of stimulus-specific computation. The recurrent input makes the framework adaptable to multi-neuron recordings \emph{in vivo}. While intracellular recordings from many neurons \emph{in vivo} are very hard to perform, the rapid development of new recording techniques (e.g. voltage-sensitive dyes) makes the future availability of subthreshold data with sufficient time-resolution at least conceivable. The full-fledged model would allow questions regarding the relative importance of background activity, recurrent activity due to computation in the circuit, and activity directly evoked by sensory stimuli to be answered in a systematic way. In this setup, the contribution of the GP-distributed membrane potential to the overall fluctuations would be reduced (since it has to capture less unrecorded neurons) while the contribution of the recorded neurons would increase. This modified model can be seen as a generalization of the stochastic spike-response model \cite{Gerstner:2002usa} or a generalization of the GLM (if the internal variable of the GLM is interpreted as the membrane potential). 

So far, we assumed that weak synaptic inputs are captured by the Gaussian process while the strong inputs that lead to the spikes are captured by the spike-related kernel $\alpha$. A straightforward extension of the model would be to consider additional intermediate inputs that cannot be captured by the GP nor by the spike-related kernel $\alpha$ but that can drive the neuron to emit (with a given probability) an action potential. Those intermediate input could be modeled as filtered Poisson events. The inclusion of those latent events would increase the complexity of the model and at the same time change some of the fitted parameters. In particular, we expect that it would increase the coupling $\beta$ between the membrane potential and the firing rate and reduce the optimal delay $\delta$ between the decision time and the peak of the action potential. This could also provide a better way to separate the subthreshold dynamics (which depends on the input activity) from the suprathreshold dynamics (which would depend only on the neuron dynamics, and not on the strong inputs that it receives, as it is the case now).

A central assumption of our model is that of a Gaussian marginal distribution of the subthreshold potential. Although it is remarkably valid for the dataset considered here (i.e. the HVC dataset $\mathcal{D}_1$ see also Fig.~\ref{Fig:4}B), datasets characterized by a distinctly non-Gaussian voltage distribution even after spike-related kernel removal are beyond the scope of the current model. In order to address this limitation, the Gaussian process could be extended to a different stochastic process, e.g. a nonlinear diffusion process, permitting non-Gaussian and in fact arbitrary marginal distributions. Moreover, a reset behavior similar to the one exhibited by an integrate-and-fire model \cite{Brette:2005ke} could be achieved with a non-stationary GP which features a mean which is reset after a spike. Both modifications would have a severe impacton the technical difficulty of model fitting. Therefore, the Gaussian assumption can be regarded as a useful compromise which is preferable over a perfect account for the skewness of the marginal distribution. 

The spike-related kernel method to separate subthreshold and suprathreshold dynamics is an important feature of the model which is used to rid the membrane potential recording of stereotypic waveforms associated with a spike. The spike related kernel as modeled in the AGAPE has no bearing on the probability of the spikes, whereas the adaptation kernel $\eta$ which modulates the firing rate after a spike is not visible in the somatic membrane potential dynamics. A simple extension of the model could include spike-triggered adaptation currents which affect both the somatic membrane potential as well as the firing intensity. Another possible extension is to allow the firing probability to depend on a filtered version of the subthreshold potential $u$ instead of the instantaneous value of $u$ at a time $\delta$ before the peak of the action potential. Both of the mentioned extensions would improve the biophysical interpretability of the AGAPE, but they would also vastly increase the number of parameters. Therefore, a model comparison would be required to determine what level of model complexity is required in order to characterize the statistics of the recording.

In the present study, the AGAPE was fit to different datasets of two different animals and brain regions. A systematic fitting to \emph{in vivo} intracellular data from a wide range of animals and brain regions would constitute a classification scheme which does not only complement existing classifications of neurons which are based on electrophysiological, morphological, histological, and biochemical data; such as the one in \cite{Markram:2004ek}, but which is in direct relationship with the computational tasks the brain is facing \emph{in vivo}.

Another application of the AGAPE could be in the context of a normative theory of short-term plasticity. Indeed, it has been recently hypothesized that short-term plasticity performs Bayesian inference of the presynaptic membrane potential based on the observed spike-timing \cite{Pfister:2010kf,Pfister:2009wi}. According to this theory, short-term plasticity properties have to match the \emph{in vivo} statistics of the presynaptic neuron. Since the AGAPE provides a realistic generative model of presynaptic activity under which inference is supposedly performed, our model can be used to make testable predictions on the dynamical properties of downstream synapses. 

\section{Materials and Methods}
\label{MM}
\subsection{Description of the datasets used}
\begin{enumerate}
\item Dataset $\mathcal{D}_1$ is a recording from a HVC neuron of an anesthetized Zebra Finch (Ondracek and Hahnloser, unpublished recordings). The recording has a total length of 270 seconds at 32 kHz (see Fig.~ \ref{Fig:1}A for a snippet of this recording) and contains 2281 action potentials. 
\item Dataset $\mathcal{D}_2$ is another recording from a projection cell in HVC of Zebra Finch, but this time the animal is awake (Vallentin and Long, unpublished recordings). It consists of 6 individual recordings which together have a length of 152.5 seconds at 40 kHz. This dataset is used for model comparison (see below).
\item Dataset $\mathcal{D}_3$ is from similar conditions as $\mathcal{D}_2$ (Vallentin and Long, unpublished recordings, see \cite{Long:2011db,Hamaguchi:2014jt, Vallentin:2015hf} for similar recordings) and has a length of 60 seconds.
\item Dataset $\mathcal{D}_4$ consists of 19 individual trials of 4.95s duration at 20 kHz. The recording was obtained from a pyramidal neuron in layer 2/3 of awake mouse visual cortex \cite{Haider:2014dc}.
\end{enumerate}

\subsection{Preprocessing}
Intracellular voltage traces are often recorded at a rate between 20 and $40\; \mathrm{kHz}$. This allows the action potentials to be resolved very clearly and precise spike timings to be extracted. However, for the study of the subthreshold regime, this high sampling rate is not required, and therefore the data may be down-sampled to roughly 1 kHz after obtaining the precise spike timings. Prior to down-sampling, we smooth with a median filter of 1ms width in order to truncate the sharp action potential peaks and avoid artifacts (see details below).

We define the spike peak times $\hat t_{i,\text{peak}}$ operationally as the time where the local maximum of the action potential is reached. This means that $\hat t_{i,\text{peak}}$ occurs after action potential onset, and hence the spike-related kernel has to extend to the past of $\hat t_{i,\text{peak}}$. The spike-related kernel starts at the nominal spike time $\hat t_i$ which is shifted from the peak time by a fixed amount $\delta$, i.e. $\hat t_{i,\text{peak}}=\hat t_{i}+\delta$. The nominal spike times $\hat t_i$ are then binned to 1 ms, yielding a binary spike train $s_i=0,1$.

For $\usom(t)$ we use a preprocessed version of the recorded trace which has been median-filtered with a width of the filter of 1 ms and then down-sampled to 1 kHz, making it the same length as the binary spike train. This procedure preserves the relevant correlation structure of the membrane potential while reducing the computational demands of fitting as much as possible. In the data we examined, the median-filtered membrane potential has a dip after $\hat t_{i,\text{peak}}$, but unless down-sampling is done carefully, this dip sometimes occurs one timestep after $\hat t_{i,\text{peak}}$ and sometimes right at $\hat t_{i,\text{peak}}$ in the downsampled $u_{\text{som}}$. Since this dip will have to be captured by the spike-related kernel which has a fixed shape for all action potentials, the down-sampling procedure has to ensure that the dip occurs always in the first time-step. We solved this problem by setting the down-sampled value of $u_{\text{som}}$ at $\hat t_{i,\text{peak}}$ (rounded to 1 ms) to the value of $u_{\text{som}}$ at $\hat t_{i,\text{peak}}$ before down-sampling.

While applying the model to the raw recording $u_{\text{raw}}$ directly (without first filtering and downsampling it) is possible in principle, it comes at a massively increased computational cost. In the interest of time required to fit the model and amount of data having to be handled, it is therefore sensible to include that pre-processing stage.

\subsection{Parametrizations and initializations} 
We already introduced the parameters $u_r$, $r_0$ and $\beta$. Additional parameters are needed to describe the autocorrelation $k(t)$, the spike-related kernel $\alpha(t)$ and the adaptation kernel $\eta(t)$.

The covariance function of the GP has to be parametrized such that it can explain the autocorrelation structure of the data. Therefore, an initial examination of the empirical autocovariance of $\usom$, i.e. 
\begin{equation}
k^{\mathrm{emp}}(j\Dt)=\frac{1}{n-j-1}\sum_{i=1}^{n-j}\parenths{\usomi{i}-\frac{1}{n-j}\sum_{k=1}^{n-j}\usomi{k}}\parenths{\usomi{i+j}-\frac{1}{n-j}\sum_{k=1}^{n-j}\usomi{k+j}},
\label{eqn:kemp}
\end{equation}
for $j=0,...,j_{\mathrm{max}}$, is done in order to determine a suitable basis. Here, we used a sum of Ornstein-Uhlenbeck (OU) kernels, i.e.
\begin{equation}
k(t)=\sum_{i=1}^{n_k}\sigma_i^2  e^{-\theta_i \abs{t}},
\label{eqn:k}
\end{equation}
where $n_k=10$ and $\theta_i=2^{-i}$ ms$^{-1}$. The autocovariance has to remain positive definite. This induces the following linear constraints:
\begin{equation}
\hat{c}_i=\sum_{j=1}^{n_k}\sigma_j^2  \hat{c}_i^{(j)}>0, \quad\forall i=1,...,n,
\end{equation}
on $\sigma_i^2$, where $\hat{c}_i^{(j)}$ are the discrete Fourier transforms of the circulant basis vectors. The optimization problem is non-concave in the subspace of $\sigma_i^2$ and multiple local maxima and saddle points can occur. Therefore, multiple initializations have to be made in order to find a potential global optimum. In general, the least-squares fit of $k(t)$ to the empirical autocovariance function \eqref{eqn:kemp} yields a good starting point for the optimization. 

The spike-rate adaptation kernel is chosen to be a linear combination of ten different alpha shapes
\begin{equation}
\eta(t)=\begin{cases}
\sum_{i=1}^{n_{\eta}}w_i\brackets{\exp(-\nu_i t)-\exp(-\omega_i t)}, & t>0,\\
0, & t\leq 0,
\end{cases}
\label{eqn:eta}
\end{equation}
where we chose $n_{\eta}=10$, $\nu_i=2\omega_i$ and $\nu_i=2^{-i}$ ms$^{-1}$. 

Since the median filter time constant is short, the voltage change around the spike can be fast, requiring flexible spike-related kernel basis. Most of this flexibility is required around $t=\delta$. Because $\delta$ is adapted, we choose a discrete parametrization which has equal flexibility from $t=0$ up to a maximum $t$. In our case, this maximum is at $t=60$ ms, and therefore our parametrization of the spike-related kernel reads
\begin{equation}
\alpha(t)=\begin{cases}
a_i, &  {\mathrm if} \quad t\in [i\Delta t,(i+1)\Delta t) \\
0, & {\mathrm else}
\end{cases}
\label{eqn:alpha}
\end{equation}
where $a_i\in\mathds{R}$, $i=1,...,60$ are the free parameters. Since the spike-related kernel fitting is concave, the large number of parameters does not lead to a dramatic increase of computational time. It also does not lead to overfitting, as is evidenced by the smoothness of the fitted kernels (see updated Figs.4,5 in the main text and the new S3-S5) and by the new model comparison results (see updated Fig.6 in the main text).

\subsection{Model validation}
We performed a factorial model comparison (see Fig.~\ref{Fig:6}) where the four factors were the presence/absence of each of the following: multiple OU components in the GP autocorrelation function (see Eq.~\eqref{eqn:k}, as opposed to only one OU kernel with variable time-constant), the spike-related kernel $\alpha$, coupling between $u$ and $s$ (through $\beta$) and adaptation $\eta$, which gives a total of 16 different models. We use the nomenclature that $\mathcal{M}_0$ is the simplest model, e.g. $\alpha=\beta=\eta=0$ and only one OU component, having only four parameters ($u_r,\theta,\sigma$ and $r_0$). A subscript $G$ (for GP) indicates that we use the multiple OU basis and any other subscript indicates that the corresponding parameter is adjustable in addition to the parameters already present in $\mathcal{M}_0$ and the parameters that are associated with the subscribed ones. E.g. $\mathcal{M}_{G\alpha}$ indicates that we use the multiple OU basis and allow a non-zero spike-related kernel and that there are now 73 parameters ($\delta$, $u_r$, $\theta_i,a_i$ for $i=1,...,60$, and $\log r_0$). The parameter $\delta$ is optimized only for the 12 out of 16 models which depend on this parameter, i.e. that have at least $\beta\neq 0$ or $\alpha\neq 0$. 

For each of the models $\mathcal{M}\in\braces{\mathcal{M}_0,...,\mathcal{M}_{G\alpha\beta \eta}}$, we performed eight-fold cross-validation \cite{Arlot:2010fl} on dataset $\mathcal{D}_2$ in order to assess the models' generalization performance. The entire dataset was cut into eight equally-sized chunks $d_j$, where $j=1,...,8$, each of length 15s ($n=15000$), and six chunks of 3s $d'_j$, $j=1,...,6$ set aside as a test set ($n'=3000$). Each model was then trained on seven out of eight chunks (treating them as independent samples) giving an optimal set of parameters $\Theta^i_j=\argmax_{\Theta} p(\braces{d_k,k\neq j}|\mathcal{M}_i,\Theta)$ and training per-bin log-likelihood $p^{\mathrm{train}}_{ij}=\tfrac{1}{7n}\log p(\braces{d_k,k\neq j}|$$\mathcal{M}_i,\Theta^i_j)$. Then the validation likelihood $p^{\mathrm{valid}}_{ij}=\tfrac{1}{n}\log p(d_j|\mathcal{M}_i,\Theta^i_{j})$ of the left-out chunk \#$j$ was evaluated. The unseen data $d'_j$ is used for a final benchmark of model performance, where the best set of parameters is selected for each model, i.e. $p^{\mathrm{test}}_{ij}=\tfrac{1}{n'}\max_{k=1,...,8}\log p(d'_j|$$\mathcal{M}_i,\Theta^i_{k})$.

\section{Acknowledgments}

We would like to thank Janie Ondracek and Richard Hahnloser in Z\"urich, Switzerland, Daniela Vallentin and Michael Long at NYU, Bilal Haider and Matteo Carandini at UCL for kindly providing the data for this study. We also thank M\'at\'e Lengyel, Christian Pozzorini and Johanni Brea for helpful discussions.

\newpage

\bibliographystyle{plos2015}
\bibliography{library}

\newpage

\begin{flushleft}
{\Huge Supplementary Text \\
\Large
\textbf{A statistical model for in vivo neuronal dynamics}
}
\label{SI:1}
\\
Simone Carlo Surace$^{1,2,\ast}$, 
Jean-Pascal Pfister$^{2}$
\\
\bf{1} Department of Physiology, University of Bern, Bern, Switzerland
\\
\bf{2} Institute of Neuroinformatics, University of Zurich and ETH Zurich, Zurich, Switzerland
\\
$\ast$ E-mail: surace@pyl.unibe.ch
\end{flushleft}

\section*{Discrete Fourier Transform}
In the following and in the main text, we denote discrete Fourier transforms of vectors of length $n$ by a hat. The Fourier transformed vector is again of length $n$ and can be formally expressed as
\begin{equation}
\hat{v}=\mathds{F}v, \quad (\mathds{F}_{n})_{ij}=e^{\tfrac{2\pi I(i-1)(j-1)}{n}}, \quad i,j=1,...,n,
\end{equation}
where $I$ denotes the imaginary unit. In practice, discrete Fourier transforms are not actually computed by matrix multiplication, but by means of a Fast Fourier Transform (FFT) algorithm. 

\section*{Circulant matrices}
In order to reduce the computational complexity of the likelihood estimation, we approximate the autocovariance matrix $K$ (which is  a Toeplitz matrix $K$) with a circulant matrix $C$. By definition a circulant matrix can be expressed as 
\begin{equation}
C_{ij}=c_{(i-j\!\!\!\!\mod n) +1}
\end{equation}
we write $C=C_n(c)$. All circulant matrices of dimension $n$ can be diagonalized by the unitary discrete Fourier transform matrix $\mathds{U}=\frac{1}{\sqrt{n}}\mathds{F}_n$:
\begin{equation}
C_n(c)=\mathds{U}_n^{\dag}\diag(\mathds{F}_nc)\mathds{U}_n=\mathds{U}_n^{\dag}\diag(\hat{c})\mathds{U}_n,
\end{equation}
where $\dag$ is the conjugate transpose. This implies that $\hat{c}$ is the vector of eigenvalues of $C$, and it is a vector with real entries. This makes calculation of inverse and determinant of $C$ extremely cheap, as is multiplication of $C^{-1}$ by a vector $x\in\mathds{R}^n$, which simplifies to
\begin{equation}
C^{-1}_n(c)x=\frac{1}{n}\mathds{F}_n^{\dag}\parenths{\frac{\hat{x}}{\hat{c}}}
\end{equation}
where the vector in brackets is the component-wise quotient of the vectors $\mathds{U}_nx$ and $\mathds{F}_nc$.

\section*{Circulant approximation}

The task is now to find a circulant matrix which is as close as possible to the covariance matrix $K$. This can be formalized as the following minimization problem:
\begin{equation}
C=\argmin_{D\; \mathrm{circulant}}{D_{\mathrm{KL}}}\parenths{\mathcal{N}(m,K)||\mathcal{N}(m,D)},\quad \forall m
\end{equation}
where $\mathcal{N}$ denotes a multivariate Gaussian with specified mean vector and covariance matrix. This problem has the unique solution
\begin{equation}
c_i=\frac{1}{n}\braces{(n-i+1)k_i+(i-1)k_{n-i+2}},\quad 1\leq i\leq n, \quad k_{n+1}\equiv 0
\label{eqn:circapprox}
\end{equation}
\textit{Proof:} The Kullback-Leibler divergence between two Gaussians is given by
\begin{equation}
D_{\mathrm{KL}}\parenths{\mathcal{N}(m,K)||\mathcal{N}(m,C)}=\frac{1}{2}\brackets{\tr(C^{-1}K)-\log\det(C^{-1}K)}-\frac{n}{2}
\end{equation}
We have
\begin{equation}
C^{-1}=\mathds{U}_n^{\dag}\diag\parenths{\frac{1}{\hat{c}}}\mathds{U}_n, \qquad \det C^{-1}=\prod_{i=1}^n \frac{1}{\hat{c}_i}
\end{equation}
and hence
\begin{equation}
D_{\mathrm{KL}}\parenths{\mathcal{N}(m,K)||\mathcal{N}(m,C)}=\frac{1}{2}\sum_{i=1}^n\brackets{\frac{(\mathds{U}_n K \mathds{U}_n^{\dag})_{ii}}{\hat{c}_i}+\log \hat{c}_i}+\mathrm{const.}
\end{equation}
where the constant does not depend on $c$. We obtain the derivative
\begin{equation}
\pd{}{c_i}D_{\mathrm{KL}}\parenths{\mathcal{N}(m,K)||\mathcal{N}(m,C)}=\frac{1}{2\hat{c}_i}\brackets{1-\frac{(\mathds{U}_n K \mathds{U}_n^{\dag})_{ii}}{\hat{c}_i}}=0
\end{equation}
and therefore, at the stationary point we have
\begin{equation}
\begin{split}
\hat{c}_i&=(\mathds{U}_n K \mathds{U}_n^{\dag})_{ii}\\
         c_i&=\frac{1}{n^2}\sum_{j,l,m=1}^n (\mathds{F}_n^{\dag})_{ij} (\mathds{F}_n)_{jl}K_{lm}(\mathds{F}_n^{\dag})_{mj} \\
             &=\frac{1}{n^2}\sum_{j,l,m=1}^n K_{lm}\exp\brackets{\frac{2\pi I}{n}\parenths{-(i-1)(j-1)+(j-1)(l-1)-(m-1)(j-1)}} \\
             &=\frac{1}{n^2}\sum_{j,l,m=1}^n k_{\abs{l-m}+1}\exp\brackets{\frac{2\pi I}{n}(j-1)(l-m+1-i)}
\end{split}
\end{equation}
The sum of roots of unity over $j$ only gives a non-zero value if the integer $q=l-m+1-i$ is a multiple of $n$. Since $q$ has a maximum of $q=n-1$ when $l=n, m=i=1$ and a minimum of $q=2-2n$ when $l=1,m=i=n$, only $q=0$ and $q=-n$ are eligible. Hence,
\begin{equation}
\begin{split}
c_i&=\frac{1}{n^2}\sum_{l,m=1}^n n k_{\abs{l-m}+1}\parenths{\delta_{0,l-m+1-i}+\delta_{-n,l-m+1-i}} \\
             &=\frac{1}{n}\sum_{r=-n+1}^{n-1} (n-\abs{r}) k_{\abs{r}+1}\parenths{\delta_{i-1,r}+\delta_{i-1,n+r}} \\
             &=\frac{1}{n}\braces{(n-i+1)k_i+(i-1)k_{n-i+2}}, \quad k_{n+1}\equiv 0
\end{split}
\end{equation}
The second equality is obtained by reparametrizing $r=l-m$. $\Box$

\section*{Sampling using FFTs}
In order generate a sample $u$ of length $n$ from a multivariate Gaussian with mean vector $m$ and circulant covariance matrix $C=C_n(c)$, one generates a zero-mean white noise vector $x$ with unit variance and then uses FFTs to compute $u$, i.e.
\begin{equation}
u=m+\frac{1}{n}\mathds{F}_n^{\dag}\brackets{\parenths{\hat{c}}^{1/2}\hat x}=m+\frac{1}{n}C_n\parenths{\mathds{F}_n^{\dag}\parenths{\hat{c}}^{1/2}}x
\end{equation}
as the following calculation shows, the covariance comes out correctly
\begin{equation}
\begin{split}
\Ev{(u-m)(u-m)^T}&=\frac{1}{n^2}C_n\parenths{\mathds{F}_n^{\dag}\parenths{\hat{c}}^{1/2}}\Ev{xx^T}C_n\parenths{\mathds{F}_n^{\dag}\parenths{\hat{c}}^{1/2}}^T\\
			&=\frac{1}{n}C_n\parenths{\mathds{F}_n^{\dag}\hat{c}}=C_n(c)=C
\end{split}
\end{equation}

\section*{Optimization method}
The optimization scheme used is a quasi-Newton method, where the vector of parameters $\Theta$ is updated according to
\begin{equation}
\Theta^{(k+1)}=\Theta^{(k)}-B^{(k)}\nabla f\parenths{\Theta^{(k)}}
\end{equation}
where $f$ is the function to be minimized (e.g. $-\log p(u_{\mathrm{som}},s)$), $\nabla f$ is the gradient, and the matrix $B$ is chosen to be
\begin{equation}
B^{(k)}=
\begin{cases}
H_f^{-1} \parenths{\Theta^{(k)}}, & \text{if }H_f \parenths{\Theta^{(k)}}\text{ positive definite} \\
\gamma^{(k)}G^{-1}, & \text{else}
\end{cases}
\label{eqn:Bmat}
\end{equation}
Where $H_f$ is the Hessian of $f$, $\gamma^{(k)}$ denotes a learning rate, and $G$ is a metric tensor on the parameter space which is used to rescale the parameters to lie in similar ranges. The learning rate $\gamma^{(k)}<0$ is increased when the previous step was successful (typically, by 10 percent), and reduced when the optimizer either runs into boundaries of the admissible parameter region or increases the value of the function (we used a reduction by a factor of 2).

Below, we derive the formulae for the gradient and Hessian required for the optimization. The derivations hold for the case where the GP covariance function $k$ is parametrized arbitrarily by $\theta_i$ and the spike-shape kernel and adaptation kernel are given by linear combinations of basis functions $\alpha^{(k)}$, $k=1,...,n_{a}$ and $\eta^{(k)}$, $k=1,...,n_{w}$ respectively.

\section*{Gradient}
The likelihood function has the form
\begin{equation}
\log p(u_{\mathrm{som}},s)=\sum_{i=1}^n\brackets{-\frac{1}{2}\log (2\pi\hat c_i)-\frac{1}{2n}\frac{\abs{\hat u_i}^2}{\hat c_i}+ s_i \log q_i+(1-s_i) \log \brackets{1-q_i}},
\end{equation}
where 
\begin{equation}
q_i=\Dt e^{\beta u_i+A_i+\log r_0}
\end{equation}
is the probability of a spike in bin $i$ and depends on all parameters except the ones that parametrize the covariance function $k$. The membrane potential is given implicitly by
\begin{equation}
u=u_{\mathrm{som}}-u_r-\alpha\ast s.
\label{eqn:u}
\end{equation}
It is worthwhile to write the derivatives of $\log p$ in the following form:
\begin{equation}
d\log p(u_{\mathrm{som}},s)=\sum_{i=1}^n\brackets{-\frac{1}{2}\parenths{\frac{1}{\hat c_i}-\frac{1}{n}\frac{\abs{\hat u_i}^2}{\hat c_i^2}}d\hat c_i-\frac{1}{n\hat c_i}\Re\braces{\hat u_i^{\ast}d\hat u_i}+\frac{s_i-q_i}{q_i(1-q_i)}dq_i},
\end{equation}
Now, let us evaluate all the terms one by one. The Fourier transform of the circulant covariance $c$ only depends on the GP kernel parameters $\theta$, i.e.
\begin{equation}
d\hat c_i=\pd{\hat c_i}{\theta_k}d\theta_k=\sum_{k=1}^{n_k}\parenths{\widehat{\pd{ c}{\theta_k}}}_id\theta_k.
\end{equation}
Let us turn to the $q$ terms next. Their differential is 
\begin{equation}
dq_i=\pd{q_i}{u_i}du_i+\pd{q_i}{A_i}dA_i+\pd{q_i}{r_0}dr_0+\pd{q_i}{\beta}d\beta=q_i(\beta du_i+dA_i+d\log r_0+u_i d\beta),
\label{eqn:dq}
\end{equation}
where by \eqref{eqn:u} and using the fact that $\alpha$ is a linear combination of basis kernels $\sum_{k=1}^{n_{\alpha}}a_k\alpha^{(k)}$
\begin{equation}
du_i=-du_r-\sum_{k=1}^{n_{\alpha}}S^{(k)}_ida_k, \quad S^{(k)}_i=(\alpha^{(k)}\ast s)_i.
\label{eqn:du}
\end{equation}
Moreover, $A_i$ is also a linear combination, so
\begin{equation}
dA_i=\sum_{k=1}^{n_{\eta}}A^{(k)}_idw_k, \quad A^{(k)}_i=(\eta^{(k)}\ast s)_i.
\end{equation}
Therefore \eqref{eqn:dq} can be written as
\begin{equation}
dq_i=q_i\parenths{-\beta du_r-\beta \sum_{k=1}^{n_{\alpha}}S^{(k)}_ida_k+\sum_{k=1}^{n_{\eta}}A^{(k)}_idw_k+d\log r_0+u_i d\beta},
\end{equation}
Lastly, by \eqref{eqn:du} we also have
\begin{equation}
d\hat u_i=-n\delta_{1i}du_r-\sum_{k=1}^{n_{\alpha}}\hat S^{(k)}_ida_k.
\end{equation}
Using
\begin{equation}
v_i=\frac{s_i-q_i}{1-q_i},
\end{equation}
all the previous results can be regrouped to yield
\begin{equation}
\begin{split}
d\log p(u_{\mathrm{som}},s)=\sum_{i=1}^n \Biggl[ -\frac{1}{2}\parenths{\frac{1}{\hat c_i}-\frac{1}{n}\frac{\abs{\hat u_i}^2}{\hat c_i^2}}\sum_{k=1}^{n_{k}}\parenths{\widehat{\pd{\hat c_i}{\theta_k}}}_id\theta_k\\
+\sum_{k=1}^{n_{\alpha}} \parenths{\frac{1}{n\hat c_i}\Re\braces{\hat u_i^{\ast}\hat S^{(k)}_i}-\beta S^{(k)}_iv_i}da_k \\
+\sum_{k=1}^{n_{\eta}} A^{(k)}_iv_idw_k  \\
+ v_id\log r_0 \\
+ u_iv_id\beta \\
+ \parenths{\frac{\delta_{1i}}{\hat c_i}\Re\braces{\hat u_i}-\beta v_i}du_r 
\Biggr],
\end{split}
\end{equation}

\section*{Hessian}
For the Hessian, we mainly need the following
\begin{equation}
-\frac{1}{2}d\parenths{\frac{1}{\hat c_i}-\frac{1}{n}\frac{\abs{\hat u_i}^2}{\hat c_i^2}}=-\frac{1}{2}\parenths{\frac{1}{ \hat c^2_i}-\frac{2}{n}\frac{\abs{\hat u_i}^2}{\hat c^3_i}}d\hat c_i+\frac{1}{n\hat c_i^2}\Re\braces{\hat u_i^{\ast}d\hat u_i},
\end{equation}
\begin{equation}
d\parenths{\frac{1}{n\hat c_i}\Re\braces{\hat u_i^{\ast}\hat S^{(k)}_i}}=-\frac{1}{n\hat c_i^2}\Re\braces{\hat u_i^{\ast}\hat S^{(k)}_i}d\hat c_i+\frac{1}{n\hat c_i}\Re\braces{\hat S^{\ast(k)}_id\hat u_i},
\end{equation}
\begin{equation}
dv_i=d\parenths{\frac{s_i-q_i}{1-q_i}}=\frac{s_i-1}{(1-q_i)^2}dq_i = \frac{\xi_i}{q_i}dq_i,
\end{equation}
where 
\begin{equation}
 \frac{\xi_i}{q_i}=\frac{s_i-1}{(1-q_i)^2}.
\end{equation}
The components of the Hessian matrix are computed as follows
\begin{align}
\spd{\log p(u_{\mathrm{som}},s)}{\theta_k}{\theta_l}&=-\frac{1}{2}\sum_{i=1}^n\braces{\spd{\hat c_i}{\theta_k}{\theta_l}\brackets{\frac{1}{ \hat c_i}-\frac{1}{n}\abs{\frac{\hat u_i}{\hat c_i}}^2}-\pd{\hat c_i}{\theta_k}\pd{\hat c_i}{\theta_l}\brackets{\frac{1}{ \hat c^2_i}-\frac{2}{n}\frac{\abs{\hat u_i}^2}{\hat c^3_i}}},\\
\spd{\log p(u_{\mathrm{som}},s)}{\theta_k}{a_l}&=-\frac{1}{n}\sum_{i=1}^n\pd{\hat c_i}{\theta_k}\Re\braces{\frac{\hat u_i^{\ast}\hat S^{(l)}_i}{\hat c_i^2}},\\
\spd{\log p(u_{\mathrm{som}},s)}{\theta_k}{u_r}&=-\pd{\hat c_1}{\theta_k}\Re\braces{\frac{\hat u_1 }{\hat c_1^2}},\\
\spd{\log p(u_{\mathrm{som}},s)}{\theta_k}{w_k}&=\spd{\log p(u_{\mathrm{som}},s)}{\theta_k}{r_0}=\spd{\log p(u_{\mathrm{som}},s)}{\theta_k}{\beta}=0,
\label{fig:hess1}
\end{align}
\begin{align}
\spd{\log p(u_{\mathrm{som}},s)}{a_k}{a_l}&=\sum_{i=1}^n\brackets{-\frac{1}{n}\Re\braces{\frac{\hat S^{\ast(k)}_i\hat S^{(l)}_i}{\hat c_i}}+\beta^2S^{(k)}_iS^{(l)}_i\xi_i},\\
\spd{\log p(u_{\mathrm{som}},s)}{a_k}{w_l}&=-\beta\sum_{i=1}^n S^{(k)}_iA^{(l)}_i\xi_i,\\
\spd{\log p(u_{\mathrm{som}},s)}{a_k}{u_r}&=-\Re\braces{\frac{\hat S^{(k)}_1}{\hat c_1}}+\beta^2 \sum_{i=1}^nS^{(k)}_i\xi_i,\\
\spd{\log p(u_{\mathrm{som}},s)}{a_k}{\log r_0}&=-\beta \sum_{i=1}^nS^{(k)}_i\xi_i,\\
\spd{\log p(u_{\mathrm{som}},s)}{a_k}{\beta}&=-\beta \sum_{i=1}^nS^{(k)}_i\xi_i u_i,
\label{fig:hess2}
\end{align}
\begin{align}
\spd{\log p(u_{\mathrm{som}},s)}{w_k}{w_l}&=\sum_{i=1}^nA^{(k)}_iA^{(l)}_i\xi_i,\\
\spd{\log p(u_{\mathrm{som}},s)}{w_k}{u_r}&=-\beta\sum_{i=1}^nA^{(k)}_i\xi_i,\\
\spd{\log p(u_{\mathrm{som}},s)}{w_k}{\log r_0}&=\sum_{i=1}^nA^{(k)}_i\xi_i,\\
\spd{\log p(u_{\mathrm{som}},s)}{w_k}{\beta}&=\sum_{i=1}^nA^{(k)}_i\xi_iu_i,
\label{fig:hess3}
\end{align}
\begin{align}
\pd{^2\log p(u_{\mathrm{som}},s)}{u_r^2}&=-\frac{n}{\hat c_1}+\beta^2\sum_{i=1}^n\xi_i,\\
\spd{\log p(u_{\mathrm{som}},s)}{u_r}{\log r_0}&=-\beta\sum_{i=1}^n\xi_i,\\
\spd{\log p(u_{\mathrm{som}},s)}{u_r}{\beta}&=-\sum_{i=1}^n\brackets{v_i+\beta u_i\xi_i},
\label{fig:hess4}
\end{align}
\begin{align}
\pd{^2\log p(u_{\mathrm{som}},s)}{(\log r_0)^2}&=\sum_{i=1}^n\xi_i,\\
\spd{\log p(u_{\mathrm{som}},s)}{\log r_0}{\beta}&=\sum_{i=1}^nu_i\xi_i,
\label{fig:hess5}
\end{align}
\begin{align}
\pd{^2\log p(u_{\mathrm{som}},s)}{\beta^2}&=\sum_{i=1}^nu_i^2\xi_i.
\label{fig:hess6}
\end{align}
The components as given by equations (33-51) are then combined to form the Hessian matrix $H_f$, which is used in Eq.~\eqref{eqn:Bmat}.

\clearpage

\begin{figure}[p]
\centering
\includegraphics[width=\textwidth]{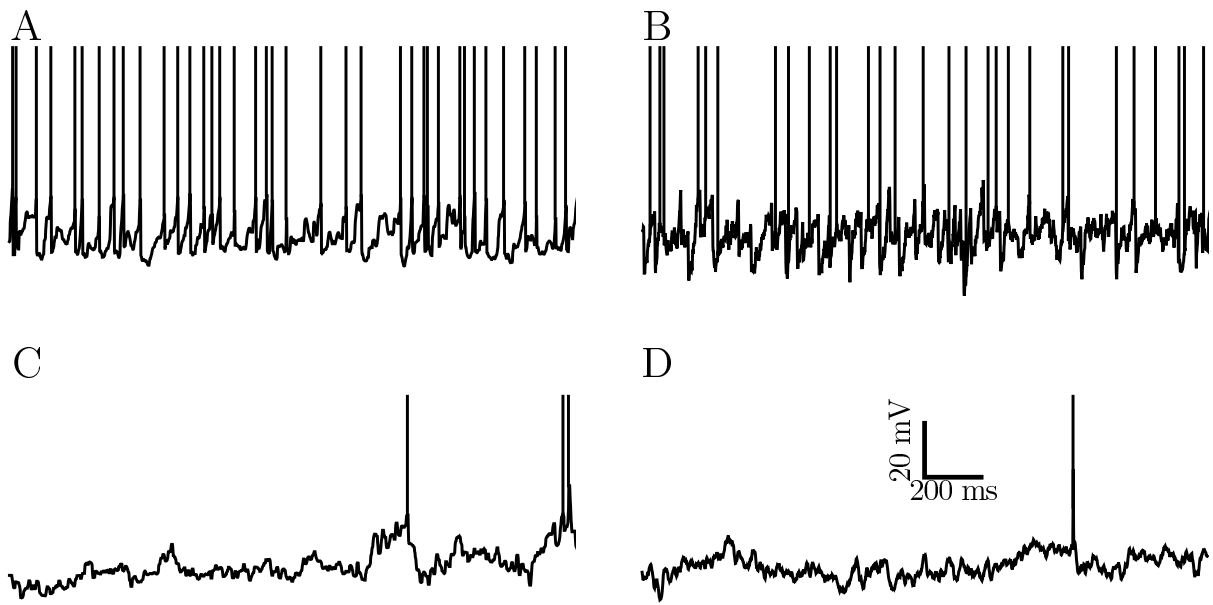}
\caption{\textbf{S2 Fig. Supplementary Figure.} \label{SI:2} Comparison of \emph{in vivo} and artificial data snippets for datasets $\mathcal{D}_3$ and $\mathcal{D}_4$, analogous to Fig.~3G,H. The scale (shown on panel D) is the same for all four panels. Vertical lines are drawn at the spiking times. (A) A 2-second sample of \emph{in vivo} activity from dataset $\mathcal{D}_3$ (Zebra Finch HVC). (B) Artificial data sampled from AGAPE with parameters learned from dataset $\mathcal{D}_3$. (C) A 2-second sample of  \emph{in vivo} activity from dataset $\mathcal{D}_4$ (mouse visual cortex). (D) Artificial data sampled from AGAPE with parameters learned from dataset $\mathcal{D}_4$. }
\label{SI:2}
\end{figure}

\begin{figure}[p]
\centering
\includegraphics[width=\textwidth]{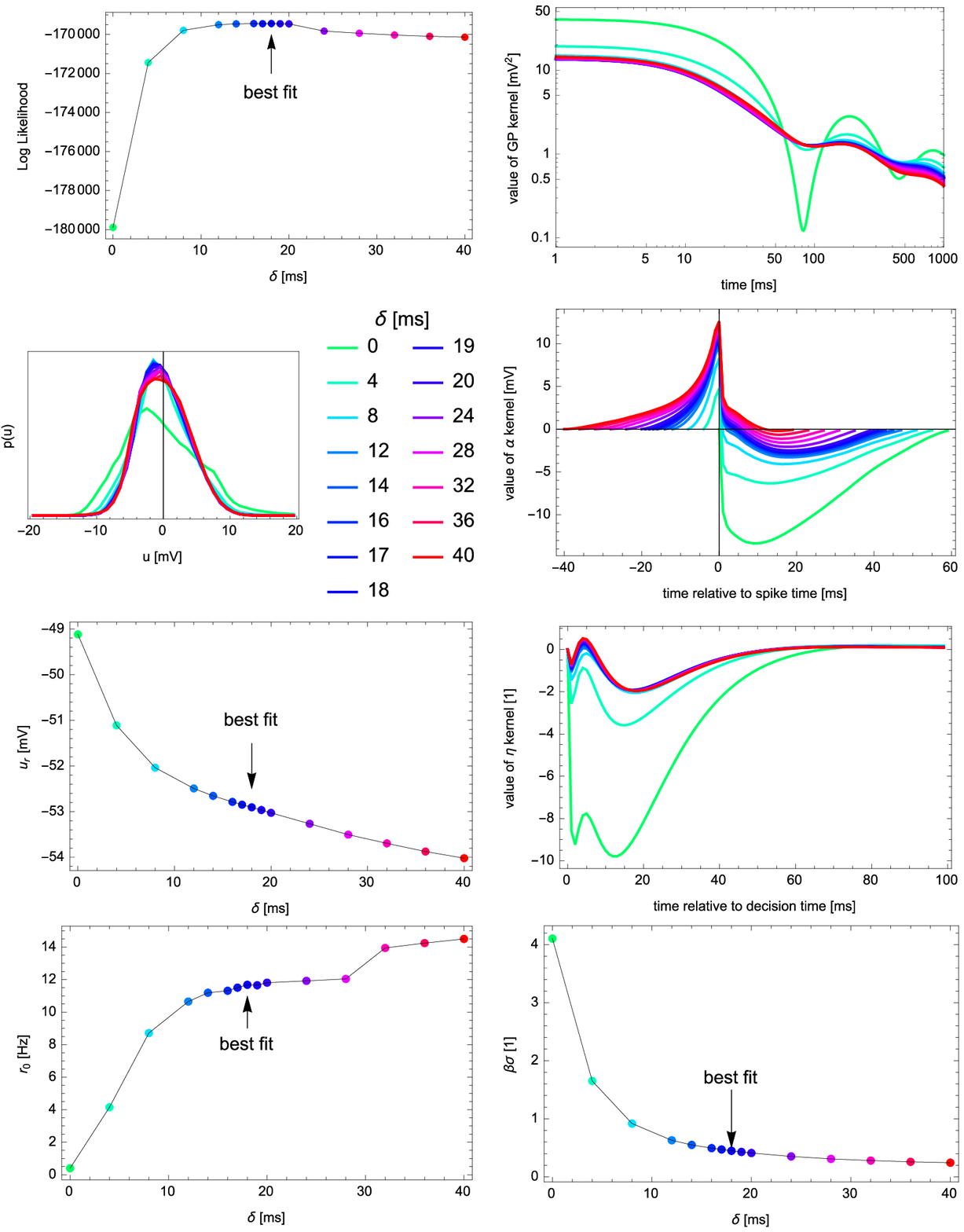}
\caption{\textbf{S3 Fig. Supplementary Figure.} \label{SI:3} The fitting result as a function of the parameter $\delta$ for dataset $\mathcal{D}_1$, see color code next to the plot of the marginal distribution of $u$ in the second row of the left column. The top left panel shows that the log likelihood peaks at $\delta=18$ ms, and the bottom right panel shows the decrease of the effective coupling stength as $\delta$ increases. }
\label{SI:3}
\end{figure}

\begin{figure}[p]
\centering
\includegraphics[width=\textwidth]{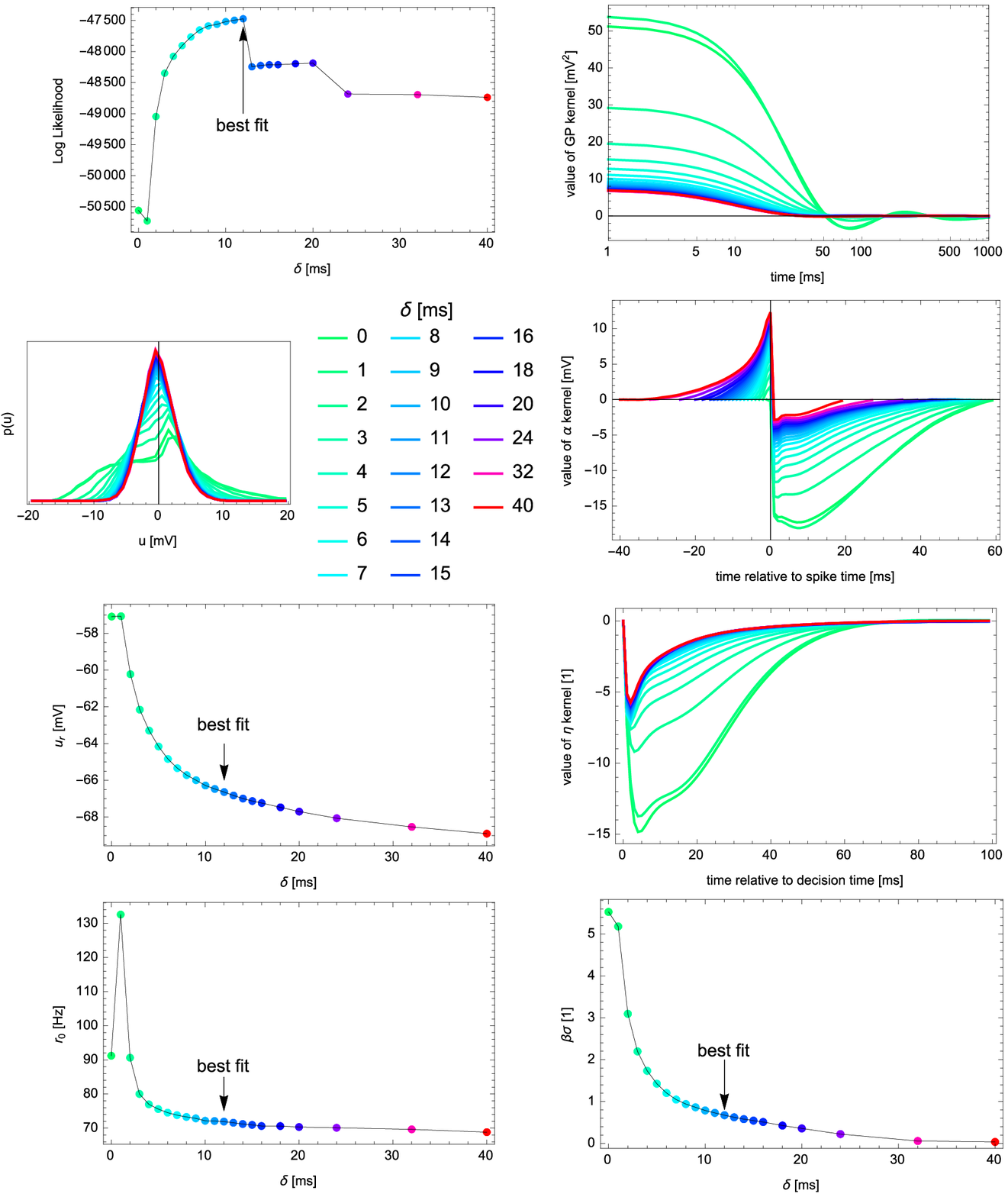}
\caption{\textbf{S4 Fig. Supplementary Figure.} \label{SI:4} The fitting result as a function of the parameter $\delta$ for dataset $\mathcal{D}_3$, see color code next to the plot of the marginal distribution of $u$ in the second row of the left column. The top left panel shows that the log likelihood peaks at $\delta=12$ ms, and the bottom right panel shows the decrease of the effective coupling stength as $\delta$ increases. }
\label{SI:4}
\end{figure}

\begin{figure}[p]
\centering
\includegraphics[width=\textwidth]{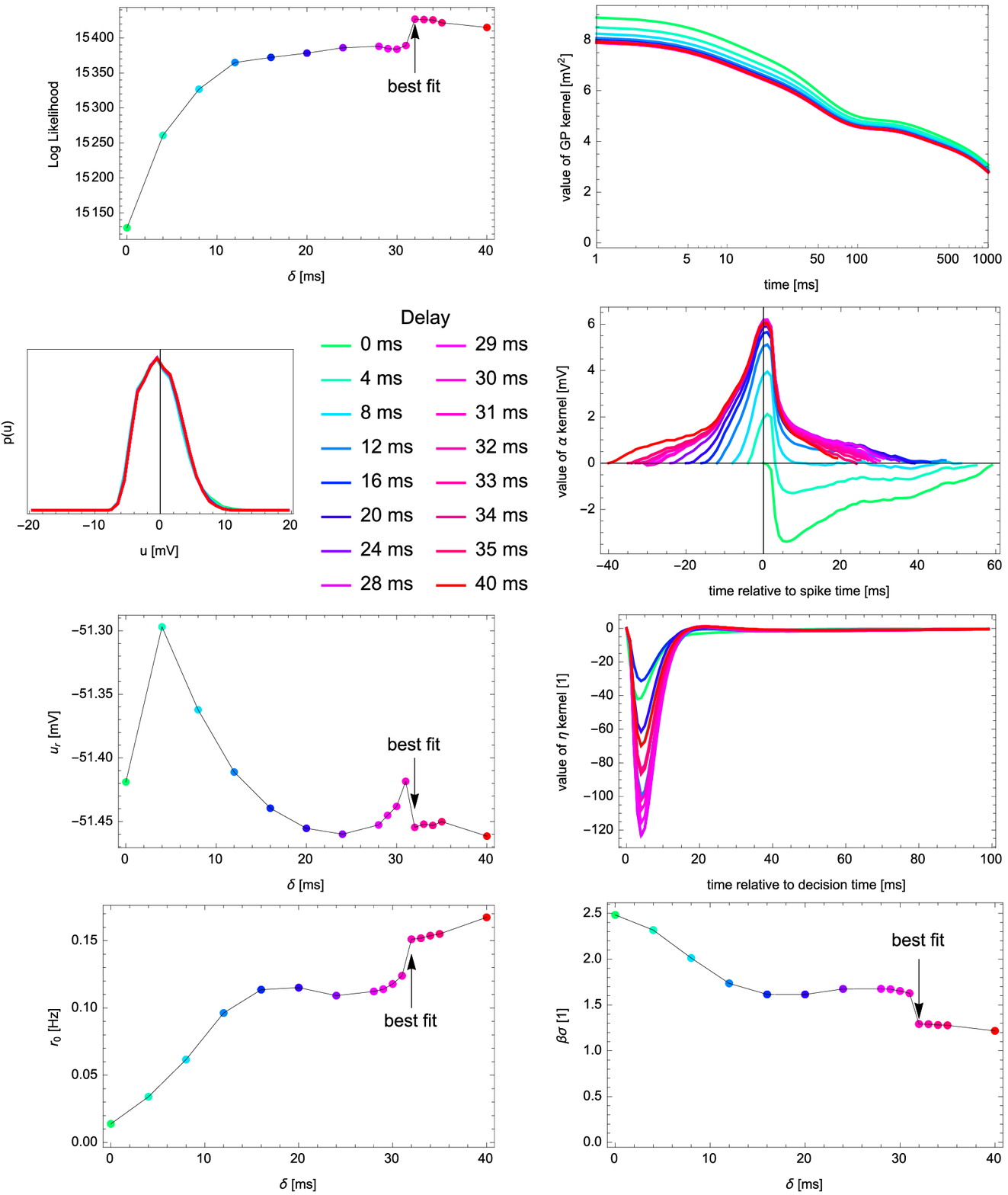}
\caption{\textbf{S5 Fig. Supplementary Figure.} \label{SI:5} The fitting result as a function of the parameter $\delta$ for dataset $\mathcal{D}_4$, see color code next to the plot of the marginal distribution of $u$ in the second row of the left column. The top left panel shows that the log likelihood peaks at $\delta=32$ ms, and the bottom right panel shows the decrease of the effective coupling stength as $\delta$ increases. }
\label{SI:5}
\end{figure}

\end{document}